\documentclass[letterpaper,11pt, margin=2.5cm]{article}
\usepackage{setspace}
\usepackage{amssymb,array}
\fontfamily{times}
\usepackage{graphicx, caption, subcaption}
\usepackage{geometry}
\usepackage{mathrsfs}
\usepackage{float}
\usepackage{subcaption}

\renewcommand{\Pr}{\mbox{$\mathbf{P}$}}
\newcommand{\Expe}{\mbox{$\mathbf{E}$}}

\begin{document}

	\begin{center}
		\textbf{Prediction of Time-to-terminal Event (TTTE)  in a Class of Joint Dynamic Models}\\
		\vspace{0.3in}
		Piaomu Liu\\
		Dept. of Mathematical Sciences\\
		Bentley University\\
		Waltham, 02451, USA
		\date{Oct 25, 2020}
	\end{center}
		
	%
		
		
	\begin{center}
		{\bf Abstract}
	\end{center}

	 In different areas of research, multiple recurrent competing risks (RCR) are often observed on the same observational unit. For instance, different types of cancer relapses are observed on the same patient and several types of component failures are observed in the same reliability system. When a terminal event (TE) such as death is also observed on the same unit, since the RCRs are generally informative about death, we develop joint dynamic models that simultaneously model the RCRs and the TE. A key interest of such joint dynamic modeling is to predict time-to-terminal event (TTTE) for new units that have not experienced the TE by the end of monitoring period. In this paper, we propose a simulation approach to predict TTTE which arises from a class of joint dynamic models of RCRs and TE. The proposed approach can be applied to problems in precision medicine and potentially many other settings. The simulation method makes \textit{personalized} predictions of TTTE and provides an empirical predictive distribution of TTTE. Predictions of the RCR occurrences beyond a possibly random monitoring time and leading up to the TE occurrence are also produced. The approach is dynamic in that each simulated occurrence of RCR increases the amount of knowledge we obtain on an observational unit which informs the simulation of TTTE. We demonstrate the approach on a synthetic dataset and evaluate predictive accuracy of the prediction method through 5-fold cross-validation using empirical Brier Score. 
\newpage	



\section{Introduction}\label{sec: Introduction}
In many settings, recurrent competing risks (RCR) such as different types of heart attacks, hospital infections, and machine component failures are common phenomena on a single observational unit. Occurrences of the RCRs are often observed with a terminal event (TE), for instance, death of a patient, and breakdown of a reliability system. When a TE happens to an observational unit, all data generation on the unit terminates. Understanding the process of time-to-terminal event (TTTE) is of high importance to research in various disciplines. 

\begin{figure}
	\centering
	\begin{subfigure}{.5\textwidth}
		\centering
		\includegraphics[width=.5\linewidth]{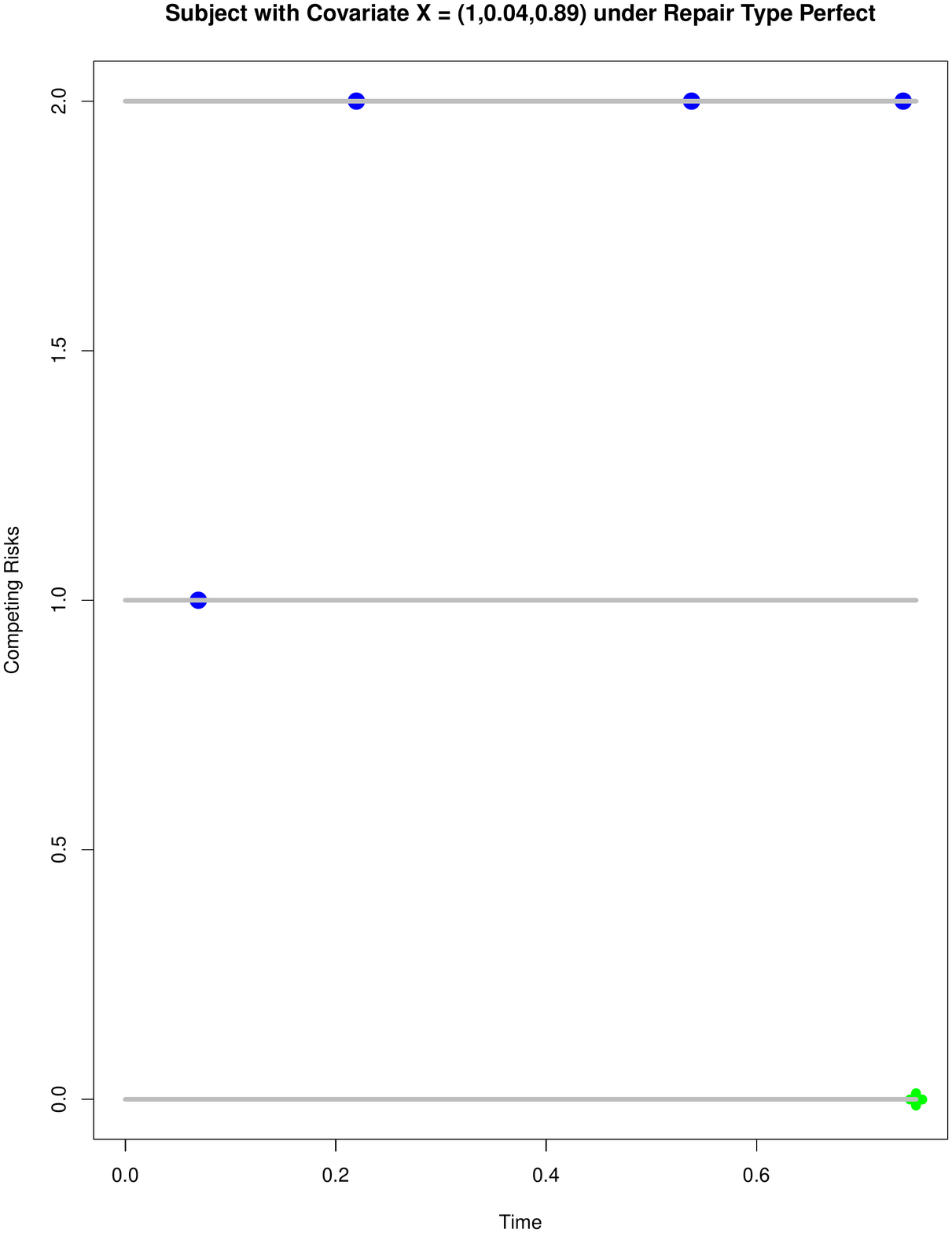}
		\caption{Synthetic Data History of a Stroke Patient}
		\label{fig:datone}
	\end{subfigure}%
	\begin{subfigure}{.5\textwidth}
		\centering
		\includegraphics[width=.5\linewidth]{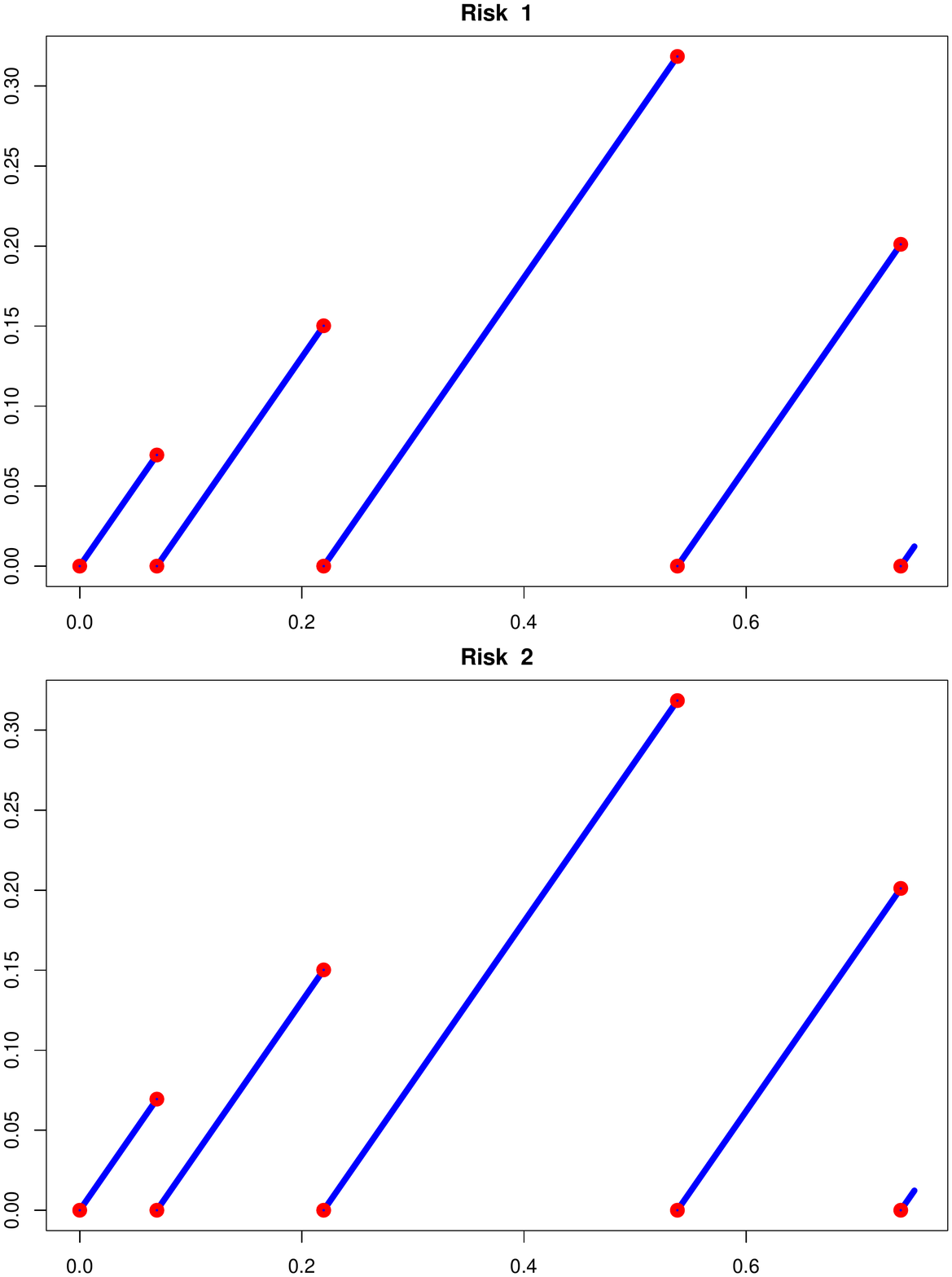}
		\caption{Effective Ages (Perfect Repair) of the Patient}
		\label{fig:datoneEAs}
	\end{subfigure}
	\caption{Synthetic Data History (\textit{left}) and Effective Ages (Perfect Repair) (\textit{Right})}
	\label{fig2:dataset2}
\end{figure}

Consider a real life scenario where event occurrences of a couple of RCRs are tracked until the end of a monitoring time period and time-independent characteristics of an observational unit are also recorded.  According to Centers of Disease Control and Prevention (CDC) and American Academy of Neurology, Transient Ischemic Attack (TIA) or mini-stroke is a warning sign of future strokes. Individuals who experience a TIA should call 911 to receive immediate medical attention in order to lower the risk of a major stroke. If TIA is ignored or no intervention is performed, one-third of the patients experience a stroke within one year of experiencing a TIA. One could experience re-occurrences of either type of strokes. If a patient of high risk to experience either TIA or a major stroke can be monitored over a period of time, and the end point of interest is death, Figure \ref{fig2:dataset2} is a visualization of one possible patient's data history. The two different types of strokes are the RCRs, and death is the TE in this scenario. Risk 1 is TIA, and Risk 2 is major stroke. TIA occurs three times, and a major stroke happens once before the end of monitoring time. For this synthetic data unit, impact of medical intervention after each recurrent event is simulated by perfect repair (\cite{pena2007semiparametric} and \cite{han2007parametric}), a common methodology to model interventions in medical research and in reliability. In this context, we use perfect repair to model the impact of a very effective medical intervention after each occurrence of the RCRs where patients recover completely. By the end of the monitoring period, the unit is still alive, which is indicated by the green cross in the left-panel of Figure \ref{fig2:dataset2}. As the RCRs are informative about the TE, to model TTTE, we build joint models (\cite{mauguen2013dynamic}, \cite{blanche2015quantifying}, and \cite{krol2017tutorial}) that simultaneously model the RCRs and the TE  in order to take into account the dependency between the processes. 
To account for the fact that the TE risk is related to event occurrences of the RCRs, the joint models are dynamic where the risk of the TE is updated as the RCRs accrue over time. In this scenario, the RCRs and TE as well as the dependency between them can be modeled on a separate sample of $n$ units of stroke patients who are similar to the observational unit. Some of the training observations experience the TE by the end of a possibly random monitoring time, and the others do not. Since the patient is still alive by the end of the monitoring time, a more significant challenge in modeling is to predict TTTE for the unit after joint modeling the RCRs and the TE successfully. 

In situations similar to the aforementioned research setting, for example, in precision medicine, the dynamic prediction question is how to predict TTTE for new units which are not used for building the joint dynamic models. In other applications where we have tracked the history of the RCRs of a new unit by some time $\tau$, including possibly other time-independent characteristics, the data are similar to  those in Figure  \ref{fig2:dataset2}. If the unit has not experienced TE by $\tau$, how to estimate the survival probability of the unit in $s$ years after $\tau$? In order to produce \textit{personalized} prediction of TTTE, how does a prediction method use information both from the training data and \textit{individual} data history of a unit to update risks of the RCRs and the TE? 
In this paper, we propose a simulation approach that provides a solution to the dynamic prediction problem of TTTE. The proposed prediction method arises from joint dynamic modeling of RCRs and TE (\cite{liu2015dynamic}) and offers researchers a tool to quantify the risk of the TE occurrence. 

\subsection{Literature Review}
In recent years, due to its significance as a prognostic tool in medical treatments, there is an increasing interest in methodological development of dynamic predictions of TTTE in medical applications of joint dynamic modeling (\cite{mauguen2013dynamic}, \cite{njagi2013joint} and \cite{proust2014joint}). In \cite{mauguen2013dynamic}, three prediction schemes with different information levels were considered to estimate the probabilities of terminal event (TE) occurring over the interval $[t, t+w)$, conditional on event history up to time $t$. Joint frailty models are proposed to model dependent recurrent event times, the dependency between multiple recurrent events and the terminal event (TE).  The frailty term is integrated out in computing the conditional probability of TE occurrence to form a marginal prediction approach. 

The dynamic aspect of joint modeling is emphasized in existing works (\cite{mauguen2013dynamic}, \cite{blanche2015quantifying}, \cite{liu2015dynamic} and \cite{krol2017tutorial}). In dynamic modeling of recurrent event data (\cite{pena2006dynamic}), interventions (\cite{han2007parametric}) are considered as they often happen after an event occurrence (\cite{montoto2002survival}, \cite{pena2006dynamic}, \cite{taylor2014nonparametric} and \cite{agustin1999dynamic}). When we jointly model RCRs and TE dynamically, the models should also consider impact of past RCR occurrences on future risks of the RCRs and the TE. When more than one recurrent event are monitored for the same unit related to the TE, assuming data are generated in continuous time so that occurrences of two different competing risks cannot happen simultaneously, existing approaches in dynamic modeling can be applied and extended to jointly model RCRs and a TE (see \cite{liu2015dynamic}). The joint models in (\cite{liu2015dynamic}) are dynamic in that interventions taking place after each RCR occurrence are considered, and impact of past RCR event occurrences are also taken into account to update the risk of TE as data accrue over time. Each additional event occurrence of the RCRs increases our knowledge about the risk of future event occurrences as well as the risk of TE. 

To tackle the dynamic prediction problem, the marginal prediction approach only provides an estimate of the TE occurrence probability. The proposed dynamic prediction method simulates many paths of TTTE and provides an empirical predictive distribution of TTTE. For a large number of simulated TTTE paths, the predicted survival probability of the unit at time $t+s, s > 0,$ is provided. To make predictions for observational units which are not used in model parameter estimation of the \textit{joint dynamic models} in \cite{liu2015dynamic}, we track data history of the units until time $t,$ which can be the end of random monitoring time $\tau.$ To update risks of the RCRs and the TE over time, the prediction method combines individual data history of a unit and model parameter estimates of the joint dynamic models on training data.  In order to simulate a single path of TTTE, the RCRs beyond time $t$ are simulated until the occurrence of TE. Each simulated RCR occurrence updates our knowledge about future risks of the RCRs and the TE. Each simulated path of TTTE together with the simulated RCRs provides a possible picture of data evolution of the observational unit beyond $t$, which is \textit{personalized}. 

To our best knowledge, no existing dynamic prediction method provides an empirical predictive distribution of TTTE. The simulated RCRs also provide additional insight into the personalized simulation of the TE occurrence. From an empirical predictive distribution, we are able to obtain summary statistics of important quantities of TTTE. From a large number of simulated TTTE paths, we are able to summarize important quantities about the RCRs as well. For instance, the average and median number of occurrences of a particular RCR between time $t$ and the TE occurrence per path. We organize our paper as the following: in Section 2, we introduce the data and relevant stochastic processes. We also describe the joint dynamic models with frailty. In Section 3, we lay out the parameter estimation procedure of the joint dynamic models, which is relevant to the proposed dynamic prediction method. In Section 4, we motivate the dynamic prediction problem of TTTE and describe the simulation algorithm. In Section 5, we describe empirical Brier score as the measure of predictive accuracy of the proposed method. In Section 6, we show model parameter estimates and demonstrate the proposed prediction method on synthetic datasets. We then present empirical brier scores at different time points of interest using 5-fold cross validation. In Section 7, we provide concluding remarks.

\section{Data and the Stochastic Processes}\label{sec: DataProcess}
In this section, we describe a synthetic training dataset of size 50 and the relevant stochastic processes. We show a sample of $n = 50$ observational units in Figure \ref{Manydat} as the training set to build the joint dynamic models (see \cite{liu2015dynamic}). We also describe the underlying stochastic processes. Let $(\Omega, \mathscr{F}, \textbf{P} )$ be some probability space. Define $\textbf{F} = \{\mathscr{F}_s | 0 \leq s \leq s^{\ast}\}$ a history or filtration on the same probability space. $\mathscr{F}_s$ contains all information at time $s$.  For a single unit $i$, the stochastic processes are
\begin{enumerate}
	\item $\{N^{\dagger}_{qi}(s): s \geq 0 \}$: counting process for the $q$th competing risk.
	\item $\{N^{\dagger}_{0i}(s): s \geq 0\}$: counting process for the terminal event.
	\item $\{Y^{\dagger}_{i}(s): s \geq 0 \}$: at-risk process.
	\item $\{\mathcal{E}_{qi}(s): s \geq 0\}$: the effective age process.
\end{enumerate}

($N^{\dagger}_{qi}(s), q = 1, 2, \cdots, Q$) and $N^{\dagger}_{0i}(s)$ are counting processes (\cite{andersen1985counting}, and \cite{andersen2012statistical}) and $Y_i^{\dagger}(s)$ are predictable processes (\cite{fleming2011counting} and \cite{kalbfleisch2011statistical})
with respect to $\mathscr{F}_{s-}$. $N^{\dagger}_{qi}(s)$ is the number of times that the $q_{th}$ recurrent event occurred over the time period $[0,s]$. $N^{\dagger}_{0i}(s)$ takes on value 1 if the TE occurred over the time inverval $[0, s]$, and 0 if the unit is still at-risk at time $s$. For unit $i$, the calendar times of the $q_{th}$ RCR occurrences are $0 \equiv S_{qi0} < S_{qi1} < \cdots < S_{qi{K_{qi} + 1}} \equiv \min(\tau_i, T_i)$, where $T_i$ is the true TTTE. $\tau_i$ is the random monitoring time independent of the RCRs and TE. $K_{qi}$ is the total number of $q_{th}$ RCR on $[0, \min(\tau_i, T_i)],$ which is also random. We  define the $\mathbb{F}$-predictable and observable processes given by $\{\mathcal{E}_{qi}(s): s \ge 0\}$ as the \textit{effective age processes},
whose paths are piecewise continuous, increasing, and differentiable in each of the random sub-intervals $(S_{qij-1},S_{qij}]$ for $j = 1,2,\ldots,K_{qi},K_{qi}+1$.  
So, for unit $i$, and $q = 1, 2, \cdots, Q$, the observables are 
\begin{eqnarray}
\textbf{D}_i(s) = \{ \big(Y_i^{\dagger}(s), N_{qi}^{\dagger}(s-), \mathcal{E}_{qi}(s), N_{0i}^{\dagger}(s-): q = 1, 2, \cdots, Q, s \geq 0 \big), X_{i}, \tau_i\}
\end{eqnarray}
$X_{i}$ is the time-independent covariate vector associated with the RCRs and the TE. We consider two conventional effective age processes with $N_{.i}^{\dagger}(v) = \sum_{q = 1}^{Q}N_{qi}^{\dagger}(v)$:
$$\mathcal{E}_{qi}^{PER}(v) = v - S_{iN_{.i}^{\dagger}(v-)};\quad \mathcal{E}_{qi}^{PAR}(v) = v - S_{iN_{qi}^{\dagger}(v-)}.$$

\begin{figure}
	\centering
	\begin{subfigure}{.5\textwidth}
		\centering
		\includegraphics[width=.5\linewidth,height = 0.15\paperheight]{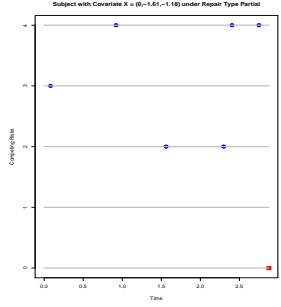}
		\caption{Data History of a Single Unit}
		\label{fig:datone}
	\end{subfigure}%
	\begin{subfigure}{.5\textwidth}
		\centering
		\includegraphics[width=.5\linewidth,height = 0.15\paperheight]{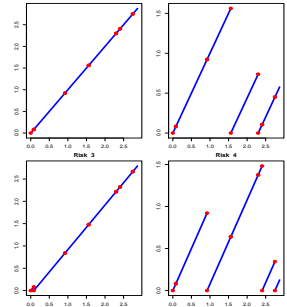}
		\caption{Effective Ages (Partial Repair) of the Unit}
		\label{fig:datoneEAs}
	\end{subfigure}
	\caption{Data History (\textit{left}) and Effective Ages (Partial Repair) (\textit{Right})}
	\label{fig:dataset1}
\end{figure}

\begin{figure}
	\centering
	\includegraphics[width=.8\linewidth,height = 0.4\paperheight]{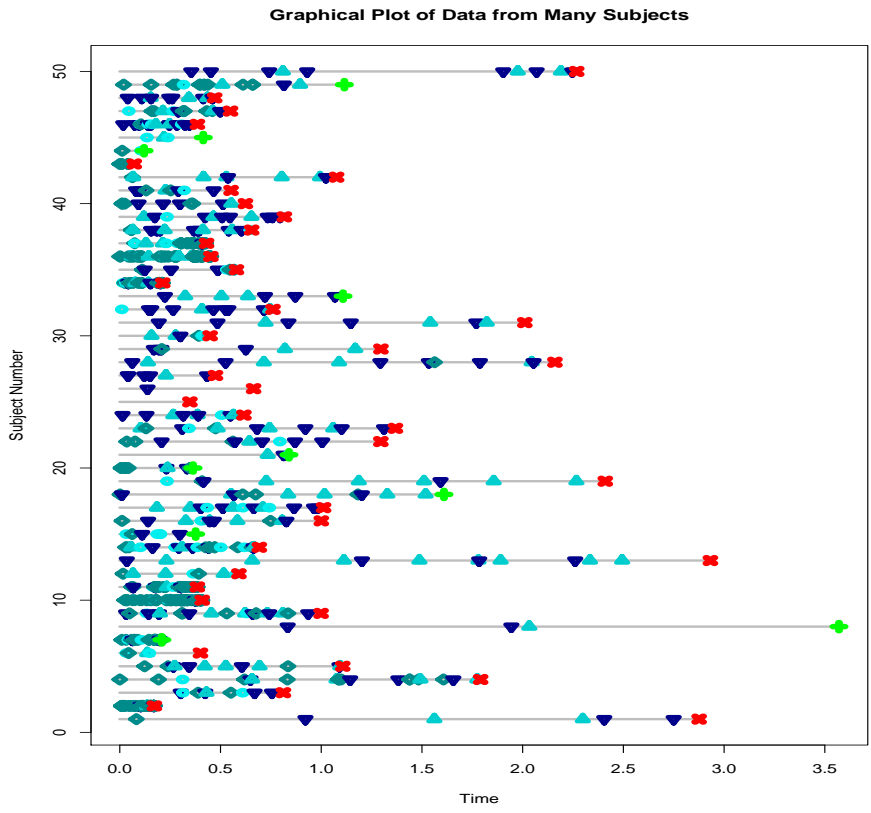}
	\caption{Many observational units in a training Set (n = 50).
	All Q=4 RCR occurrences are plotted using colored symbols on the same line for each observational unit. 
	A red cross indicates the TE occurrence of a unit. Green crosses indicate at-risk units by the end of monitoring time. 
	}
	\label{Manydat}
	\end{figure}

$\mathcal{E}_{qi}^{PER}(v)$ models an intervention after a recurrent event that resets all RCR effective ages to 0 at time $v$, which models all risks to start at effective age 0 after the intervention (\cite{pena2007semiparametric} and \cite{han2007parametric}). $\mathcal{E}_{qi}^{PAR}(v)$ models a scenario where an intervention only happens to the $q_{th}$ risk that experiences the most recent event prior to $v$. Since no intervention is performed to the other risks, these risks do not reset their effective ages. In Figure \ref{fig:dataset1}, we show an example of a single unit under partial repair. The unit experiences TE by the end of monitoring period $\tau$, and the occurrence of TE is indicated by a red cross. Let $T$ denote TTTE. The observed time to TE is $T',$ where $T' = \min(T, \tau)$. In this example, $T' = 3.$ On the right-hand panel, we plot effective ages under partial repair. Effective ages of all Q = 4 risks are plotted. Blue lines represent effective ages over time, and red solid circles indicate a RCR occurrence. 


\subsection{Model Description}\label{sec: ModelDescri}
To obtain parameter estimates from training data using the joint dynamic models in \cite{liu2015dynamic}, we describe the cumulative intensity process of risks $q, q = 1, 2, \cdots, Q,$ 
\begin{eqnarray}\label{eq: RCR}
A^{\dagger}_{qi}(s|Z_i) &=& Z_i\int_{0}^{s}Y_i^{\dagger}(v)\rho_{q}(\textbf{N}^{\dagger}_{i}(v-);\alpha_q)\exp(X_i\beta_{q})
\lambda_{q0}(\mathcal{E}_{qi}(v))dv
\end{eqnarray}
where $Z_i$ is the frailty, assuming $Z_i \stackrel{iid}{\sim} Ga(\xi, \xi), \xi > 0$. $Ga(\xi, \xi)$ is a Gamma random variable with parameter $\xi$. This term is introduced to model other unobserved factors contributing to the association between the RCRs and TE, as well as the correlation between inter-event times of the $q_{th}$ RCR. The cumulative intensity process of TE is 
 \begin{eqnarray}\label{eq: TE}
A^{\dagger}_{0i}(s|Z_i) &=& Z_i\int_{0}^{s}Y_i^{\dagger}(v)\rho_{0}(\textbf{N}^{\dagger}_{i}(v-);\gamma)\exp(X_i\beta_{0})
\lambda_{0}(v)dv
 \end{eqnarray}
The $\rho_q(.; \alpha_q)$ and $\rho_0(.;\gamma)$ are linear combination of recurrent competing event counts by time $v$. 
$$\rho_q(\textbf{N}^{\dagger}_{i}(v-);\alpha_q) = \textbf{N}^{\dagger}_{i}(v-)^{T}\alpha_q
; \quad \rho_0(\textbf{N}^{\dagger}_{i}(v-);\gamma) = \textbf{N}^{\dagger}_{i}(v-)^{T}\gamma$$

where $\alpha_q = (\alpha_qI\{q = q'\}), q' = 1, 2, \cdots, Q.$ So, $\alpha_q$ is a $Q \times 1$ vector of 0 or 1 elements, where all elements are 0 except for the $q_{th}$ position. For both the RCRs and the TE, the functions capturing effects of past event occurrences on the instantaneous probability of event occurrence conditional on history. We choose the $\rho_q(\textbf{N}^{\dagger}_{i}(v-);\alpha_q)$ and $\rho_0(\textbf{N}^{\dagger}_{i}(v-);\gamma)$ as linear combinations of the $N_{qi}^{\dagger}(v-), q = 1, 2, \cdots, Q,$ to avoid explosion (\cite{gjessing2010recurrent}).

Let $\Theta = \{\xi,\gamma,\beta_0, (\alpha_q,\beta_q ; q = 1,2,\cdots,Q)\}
\cup \{(\Lambda_{q0}(s), q = 1, 2, \cdots, Q), \Lambda_{0}(s) | 0 \leq s \leq s^{\ast}\}$ denote the vector of unknown parameters in the joint dynamic models. $\hat{\Theta}$ is the vector of model parameter estimates. 
\section{Model Estimation}\label{sec:ModelEst}
Given frailty $Z_i$, the likelihood on the $i$th unit is
\begin{eqnarray*}\label{likelihoodFrail} 
	\mathscr{L}_c (s, \Theta|Z_i) &\equiv &\prod_{v=0}^{s} \Pr\{\bigcap_{q=1}^{Q}[dN^{\dagger}_{qi}(v) = dn_{qi}(v)];[dN^{\dagger}_{0i}(v)=dn_{0i}(v)]|\mathscr{F}_{v-},Z_i\}\\
	&= &\{\prod_{q=1}^{Q}[dA_{qi}(v|Z_i)]^{dn_{qi}(v)}[1 - dA_{qi}(v|Z_i)]^{1 - dn_{qi}(v)}\} \\
	&\times&\{[dA_{0i}(v|Z_i)]^{dn_{0i}(v)}[1-dA_{0i}(v|Z_i)]^{1 - dn_{0i}(v)}\}	  
\end{eqnarray*}
where $dn_{qi}(v), dn_{0i}(v) \in \{0, 1\}$ and $\sum_{q = 1}^{Q}dn_{qi}(v) + dn_{0i}(v) \leq 1$. 

\subsection{Generalized At-risk Processes}
Effective age processes in the RCR submodels are incorporated into baselines of the RCR intensity processes. We use the doubly-indexed process (see \cite{pena2001nonparametric}, and \cite{pena2007semiparametric}) to derive a generalized-at-risk process for $q = 1, 2, \cdots, Q$. The doubly-indexed process is $Z_{qi}(s,t) = I\{\mathcal{E}_{qi}(s) \leq t\}$, where $s$ is the calendar time. $Z_{qi}(s,t)$ is to be distinguished from the frailty variable $Z_i$ in the model and takes on value 1 if the effective age of unit $i$ is less than or equal to $t$ at calendar time $s$. To develop an inference procedure, we first consider a compensator of $N_{qi}^{\dagger}(t)$, the $q_{th}$ counting process without frailty $Z_i$. Equation (\ref{eq: RCR}) becomes:
\begin{eqnarray}\label{eq: RCRNoFrail}
A^{\dagger}_{qi}(s) &=& \int_{0}^{s}Y_i^{\dagger}(v)\rho_{q}(\textbf{N}^{\dagger}_{i}(v-);\alpha_q)\exp(X_i\beta_{q})
\lambda_{q0}(\mathcal{E}_{qi}(v))dv
\end{eqnarray}

Then, the doubly-indexed processes for the $q_{th}$ RCR without frailty $Z_i$: 
\begin{eqnarray*}
	N_{qi}(s,t) = \int_{0}^{s}Z_{qi}(s,t) N_{qi}^{\dagger}(dv); \quad 
	A_{qi}(s,t) = \int_{0}^{s}Z_{qi}(s,t) A_{qi}^{\dagger}(dv)\\
	M_{qi}(s,t) = N_{qi}(s,t) - A_{qi}(s,t) = \int_{0}^{s}Z_{qi}(s,t) M_{qi}^{\dagger}(dv)
\end{eqnarray*}

For fixed $s$, $M_{qi}(s,.)$ is not a martingale but has mean 0 (see \cite{pena2001nonparametric}, and \cite{pena2007semiparametric}). For $q_{th}$ RCR, we obtain
$$\sum_{i = 1}^{n}M_{qi}(s, dw) = \sum_{i = 1}^{n}N_{qi}(s, dw) - S_{q0}(s,w)\Lambda_{q0}(s, dw) $$
where $S_{q0}(s,w) $ is the \textit{aggregate at-risk process} of the $q_{th}$ risk, and 
$$S_{q0}(s,w) = \sum_{i = q}^{n} Y_{qi}(s, dw|\alpha_a, \beta_q) =\sum_{i=1}^{n}\sum_{j=1}^{N^{\dagger}_{qi}[(s\wedge\tau_i)-]+1}I[w \in \big(\mathcal{E}_{qi}(s_{ij-1}), \mathcal{E}_{qi}(s_{ij})\big)]
\frac{\kappa_{qi}(\mathcal{E}_{qij}^{-1}(w))}{\mathcal{E}^{'}_{qi}(\mathcal{E}^{-1}_{qij}(w))} $$
where $\kappa_{qi}(\mathcal{E}^{-1}_{qij}(w)) = \rho_{q}(N^{\dagger}_{qi}(\mathcal{E}^{-1}_{qij}(w)-);\alpha_q)\exp(X_i^{T}\beta_q)$.
Adding the frailty term, we obtain the aggregated effective age processes for all RCRs as follows: for $q = 1, 2, \cdots, Q$,
\begin{eqnarray*}
	S_{q0}(s,w|\alpha_q,\beta_q,\textbf{z})& = & \sum_{i=1}^{n}Y_{qi}(s,w|\alpha_q,\beta_q,z_i)\\
	&= &\sum_{i=1}^{n}\sum_{j=1}^{N^{\dagger}_{qi}[(s\wedge\tau_i)-]+1}z_i Y_{qi}(s, dw|\alpha_q, \beta_q).
\end{eqnarray*}
 For TE process,  $$S_0(v|\gamma, \beta_0,\textbf{z}) = \sum_{i=1}^{n}Y^{\dagger}_i(v|\gamma,\beta_0,z_i) 
= \sum_{i=1}^{n}z_i I[ (\tau_i \wedge S_i) \geq v]\kappa_{0i}(v)$$ 
where $\kappa_{0i}(v) = \rho_0(\textbf{N}^{\dagger}_i(v-);\gamma)\exp(X_i^{T}\beta_0)$ with $S_i$ being time-to-terminal event.

Following the approach in  \cite{pena2001nonparametric}, given values of the finite-dimensional parameters and frailty $\textbf{Z} = \textbf{z}$, 
we estimate baseline hazards of the RCRs and TE with frailty using the expressions below
\begin{equation}\label{eq: frailLambdaEst}
\hat{\Lambda}_{q0}(s, t|\textbf{z},\alpha_q,\beta_q) = \int_{0}^{t} \frac{\sum_{i=1}^{n}N_{qi}(s,dw)}{S_{q0}(s,w|\textbf{z},\alpha_q,\beta_q)};\qquad
\hat{\Lambda}_{0}(t|\textbf{z},\gamma,
\beta_0) = \int_{0}^{t} \frac{\sum_{i=1}^{n}N^{\dagger}_{0i}(dv)}{S_{0}(v|\textbf{z},\gamma,\beta_0)}.
\end{equation}
Plugging in the estimates of the finite-dimensional parameters, the PLEs of the baseline survival functions of the RCRs ($q = 1,2,\cdots, Q$) 
and TE processes, conditional on $\textbf{Z} = \textbf{z}$, are 
\begin{eqnarray*}
	\hat{\bar{F}}_{q0}(s,t|\textbf{z}) = \prod_{w = 0}^{t}[1 - \hat{\Lambda}_{q0}(s,dw|\textbf{z})]; \qquad \hat{\bar{F}}_{0}(t|\textbf{z}) = \prod_{w = 0}^{t}[1 - \hat{\Lambda}_{0}(dw|\textbf{z})].
\end{eqnarray*}

\subsection{An EM Algorithm}\label{sec: EM}
 We develop an EM algorithm (cf. \cite{dempster1977maximum}) to estimate the finite-dimensional parameters since the frailty $Z$ is latent and hence, unobserved. Assuming $\textbf{Z} = \textbf{z}$ 
is known, we obtain the full likelihood process as below\\

$\mathscr{L}^{\dagger}[s^{\ast}| \Theta, \textbf{Z} = \textbf{z},\textbf{D}(s^{\ast})]  = \prod_{i = 1}^{n}\Big\{\frac{\xi^{\xi}}{\Gamma(\xi)}z_i^{\xi - 1}\exp(-\xi z_i)$
\begin{eqnarray*}\label{fullLH}
	&\times & \prod_{v=0}^{s^{\ast}}\prod_{q = 1}^{Q}\Big(z_iY^{\dagger}_i(v)\lambda_{q0}(\mathcal{E}_{qi}(v))
	\rho_q\Big[N_{qi}^{\dagger}(v-);\alpha_q\Big]\exp(X_i^{T}\beta_q)\Big)^{N^{\dagger}_{qi}(dv)}\\
	&\times& \exp\Big(-\int_{0}^{s^{\ast}}z_iY_i^{\dagger}(v)\lambda_{q0}(\mathcal{E}_{qi}(v))\rho_{q}\Big[N_{qi}^{\dagger}(v-);\alpha_q \Big]
	\exp(X_i^{T}\beta_q)dv\Big)\\
	&\times& \Big(z_iY^{\dagger}_i(v)\lambda_0(v)\rho_0[\textbf{N}^{\dagger}_i(v-);\gamma]\exp(X_i^{T}\beta_0)\Big)^{N^{\dagger}_i(dv)}\\
	&\times&\exp\Big(-\int_{0}^{s^{\ast}}z_iY_i^{\dagger}(v)\lambda_{0}(v)\rho_0\Big[\textbf{N}_{i}^{\dagger}(v-);\gamma \Big]\exp(X_i^{T}\beta_0)dv\Big)
	\Big\}
\end{eqnarray*}	
To compute conditional distribution of $Z_i, i = 1, 2, \cdots, n$, we use the fact that
$$Z|\Theta,\textbf{D}(s^{\ast}) \propto \mathscr{L}^{\dagger}[s^{\ast}| \Theta, \textbf{Z} = \textbf{z},\textbf{D}(s^{\ast})] \prod_{i = 1}^{n}f(Z_i|\xi).$$
We then obtain $Z_i | \textbf{D}(s^{\ast}), \Theta \stackrel{iid}{\sim} Ga(\alpha, \beta)$, with 
\begin{eqnarray*}
	\alpha(s^{\ast}) &=& \xi + \sum_{q=1}^{Q}N^{\dagger}_{qi}(s^{\ast}-) + N^{\dagger}_{0i}(s^{\ast}-)\\
	\beta(s^{\ast}) &=&  \xi + \sum_{q=1}^{Q}\int_{0}^{s^{\ast}}A^{\dagger}_{qi}(dv) + \int_{0}^{s^{\ast}}A^{\dagger}_{0i}(dv)
\end{eqnarray*}
For $i = 1, 2, \cdots, n$, the conditional expectation of $Z_i |\textbf{D}(t), \Theta$ is 

\begin{eqnarray*}
	\Expe[Z_i|\mathscr{F}_{t-},X_i,\Theta]&=& \frac{\alpha(t)}{\beta(t)} = \frac{\xi + \sum_{q=1}^{Q}N^{\dagger}_{qi}(t-) + N^{\dagger}_{0i}(t-)}
	{\xi + \sum_{q=1}^{Q}\int_{0}^{t}A^{\dagger}_{qi}(dv) + \int_{0}^{t}A^{\dagger}_{0i}(dv)}\\
	\Expe[\log(Z_i)|\mathscr{F}_{t-},X_i,\Theta]&= & \textbf{D}\textbf{G}(\xi + \sum_{q=1}^{Q}N^{\dagger}_{qi}(t-)+N^{\dagger}_{0i}(t-))
\end{eqnarray*}
$$+ \log[\Expe\{Z_i|\mathscr{F}_{t-},X_i,\Theta\}] - \log(\xi +\sum_{q=1}^{Q}\int_{0}^{t}A^{\dagger}_{qi}(dv) + \int_{0}^{t}A^{\dagger}_{0i}(dv) )$$
where $\textbf{D}\textbf{G}(\alpha) = \frac{d}{d\alpha}\log \Gamma(\alpha). $\\

The EM algorithm is described as follows:\\

\textbf{E-step}: Obtain conditional expectation of the full log-likelihood with respect  to $\textbf{Z}|(\textbf{D}(t-), \Theta)$. Let
$$\widehat{Z_i} =\Expe[Z_i|\mathscr{F}_{t-},X_i,\Theta]; \qquad \widehat{logZ_i} = \log (\Expe[Z_i|\mathscr{F}_{t-},X_i,\Theta])$$
\begin{eqnarray*}
	\Expe (\log\mathscr{L}_c^{\dagger}[s^{\ast}| \Theta, \textbf{Z}, \textbf{D}(s^{\ast}))  &=&  n \xi\log\xi - n\log\Gamma(\xi)  \\
	& + &\sum_{i = 1}^{n} \widehat{logZ_i} (\sum_{q = 1}^{Q} N^{\dagger}_{qi}(s^{\ast}) + N^{\dagger}_{0i}(s^{\ast}) + \xi - 1)\\ 
	&-& \sum_{i = 1}^{n} \widehat{Z_i} \big(\xi + \int_{0}^{s^{\ast}}[\sum_{q = 1}^{Q}A^{\dagger}_{qi}(dv) + A^{\dagger}_{0i}(dv)]\big)\\
	&+& \sum_{i = 1}^{n} \big(\sum_{q = 1}^{Q} \int_{0}^{s^{\ast}} \log a^{\dagger}_{qi}(v)N^{\dagger}_{qi}(dv) +
	\int_{0}^{s^{\ast}} \log a^{\dagger}(v)N^{\dagger}_{0i}(dv) \big)
\end{eqnarray*}

\textbf{M-step}: When values of the finite-dimensional parameters in $\Theta$ and the frailty variables $\textbf{Z}$ are given, 
baseline hazards of the RCRs and the TE can be estimated non-parametrically
as in equation (\ref{eq: frailLambdaEst}). We obtain the estimating equations below to estimate the finite-dimensional parameters.
For $\alpha_q$ and $\beta_q, q = 1, 2, \cdots, Q:$
\begin{eqnarray*}	
	&	\sum_{i=1}^{n}\int_{0}^{\tau_i}
	\Big[\frac{\frac{\partial}{\partial \alpha_q}\rho_{q}(N^{\dagger}_{qi}(v-);\alpha_q))}{\rho_{q}(N^{\dagger}_{qi}(v-);\alpha_q))}
	- \frac{\frac{\partial}{\partial \alpha_q}S_{q0}(s,\mathcal{E}_{qi}(v)|\alpha_q, \beta_q,\textbf{z})}{S_{q0}(s,\mathcal{E}_{qi}(v)|\alpha_q, \beta_q,\textbf{z})} \Big] N^{\dagger}_{qi}(dv) = 0; &\\
	& \sum_{i=1}^{n}\int_{0}^{\tau_i}
	\Big[X_i
	- \frac{\frac{\partial}{\partial \beta_q}
		S_{q0}(s,\mathcal{E}_{qi}(v)|\alpha_q, \beta_q,\textbf{z})}{S_{q0}(s,\mathcal{E}_{qi}(v)|\alpha_q, \beta_q,\textbf{z})} \Big] N^{\dagger}_{qi}(dv) = 0. &	
\end{eqnarray*}
For $\gamma$ and $\beta_0$:
\begin{eqnarray*}
	&	\sum_{i=1}^{n}\int_{0}^{\tau_i}\Big[\frac{\frac{\partial}{\partial \gamma}\rho_{0}(N^{\dagger}_i(v-);\gamma)}{\rho_{0}(N^{\dagger}_i(v-);\gamma)}
	-\frac{\frac{\partial}{\partial \gamma} S_{0}(v|\gamma, \beta_0,\textbf{z})}{S_{0}(v|\gamma, \beta_0,\textbf{z})} \Big] N^{\dagger}_{0i}(dv) = 0;\\
	&	\sum_{i=1}^{n}\int_{0}^{\tau_i}
	\Big[X_i
	- \frac{\frac{\partial}{\partial \beta_0} S_{0}(v|\gamma, \beta_0,\textbf{z})}{S_{0}(v|\gamma, \beta_0,\textbf{z})} \Big] N^{\dagger}_{0i}(dv) = 0.
\end{eqnarray*}
We follow the algorithm described below to estimate all parameters:
\begin{enumerate}
	\itemsep-0.5em 
	\item Initialize $\hat{\textbf{Z}}^{(0)} = \textbf{1}_{n \times 1}, \hat{\alpha}_q^{(0)},\hat{\beta}_q^{(0)},
	\hat{\gamma}^{(0)},\hat{\beta}_0^{(0)}$ and $\hat{\xi}^{(0)}$.
	\item Obtain $\hat{\Lambda}_{q0}^{(0)}(.), q = 1,2 \cdots,Q$ and $\hat{\Lambda}_0^{(0)}(.)$.
	\item Update to $\hat{\textbf{Z}}^{(1)}$ using $\{\hat{\Lambda}_{q0}(.),\hat{\alpha}_q^{(0)},\hat{\beta}_q^{(0)},
	q = 1, 2, \cdots,Q; \hat{\gamma}^{(0)},\hat{\beta}_0^{(0)},\hat{\Lambda}_0(.), \hat{\xi}^{(0)}\}. $
	\item Update $\hat{\xi}^{(0)}$ to $\hat{\xi}^{(1)}$.  Define $\mathscr{L}_{\xi}[s^{\ast}|\Theta, \textbf{D}(s^{\ast})] =  \Expe (\log\mathscr{L}_c^{\dagger}[s^{\ast}| \Theta, \textbf{Z}, \textbf{D}(s^{\ast}))$
	as in the $\textbf{E}$ step. $\hat{\xi}  =  \arg\max_{(\xi)}\mathscr{L}_\xi [s^{\ast}|\Theta^{(0)}, \textbf{D}(s^{\ast})]$. 
	\item With $\hat{\textbf{Z}}^{(1)}$, we update $\hat{\alpha}_q^{(0)},\hat{\beta}_q^{(0)}, \hat{\gamma}^{(0)},\hat{\beta}_0^{(0)}$ to
	$\hat{\alpha}_q^{(1)},\hat{\beta}_q^{(1)}, \hat{\gamma}^{(1)},\hat{\beta}_0^{(1)}$.
	\item Reset $\hat{\textbf{Z}}^{(1)}$ to $\hat{\textbf{Z}}^{(0)}$,
	$\hat{\alpha}_q^{(1)},\hat{\beta}_q^{(1)},
	\hat{\gamma}^{(1)},\hat{\beta}_0^{(1)}$ to $\hat{\alpha}_q^{(0)},\hat{\beta}_q^{(0)},
	\hat{\gamma}^{(0)},\hat{\beta}_0^{(0)}$.
\end{enumerate}
Repeat steps 2 - 5 until $|(\textbf{Z}^{(0)},\Theta^{(0)}) - (
\textbf{Z}^{(1)},\Theta^{(1)})| < $ tol. For example, tol = $10^{-7}$. The convergence criterion only applies to the finite-dimensional parameters in $\Theta$.
\section{Predicting Time-to-terminal-event}\label{sec3: Methods}
 With parameter estimates $\hat{\Theta}$ of the joint dynamic models described in Section \ref{sec: ModelDescri} and Section \ref{sec:ModelEst}, our proposed simulation approach dynamically predicts occurrences of the RCRs and the TE for units that are still at-risk by a pre-selected time $t$: 1) each predicted $q_{th}$ RCR occurrence updates the instantaneous probability of next event occurrence for all risks and that of the TE, conditional on history; 2) RCR occurrences are generated until the occurrence of the simulated TE; 3) we dynamically simulate an empirical predictive distribution of TTTE ; 4) we compute the predicted survival probability of an observational unit beyond $t$ at the end of some time window say, $(t, t+s], s > 0$.  A major contribution our approach makes to existing methodology is that we provide an empirical predictive distribution of TTTE as well as simulations of possible RCR occurrences leading up to the occurrence of the TE. In particular, the method is dynamic in that each simulated RCR increases the knowledge we have about the unit in order to simulate future RCR occurrences and the TE. So, individual history for each unit that we use to predict TTTE does not stop at pre-selected time $t$, for instance, at $\tau_i$, the end of the random monitoring time. The method combines individual history, and information from training data (see Figure  \ref{Manydat} in Section \ref{sec: DataProcess}) to predict TTTE dynamically. 
 
\subsection{Frailty Z Estimates}\label{sec: EstZ}
%

\begin{figure}
	\centering
	\begin{subfigure}{.5\textwidth}
		\centering
		\includegraphics[width=.8\linewidth, height = 0.25\paperheight]{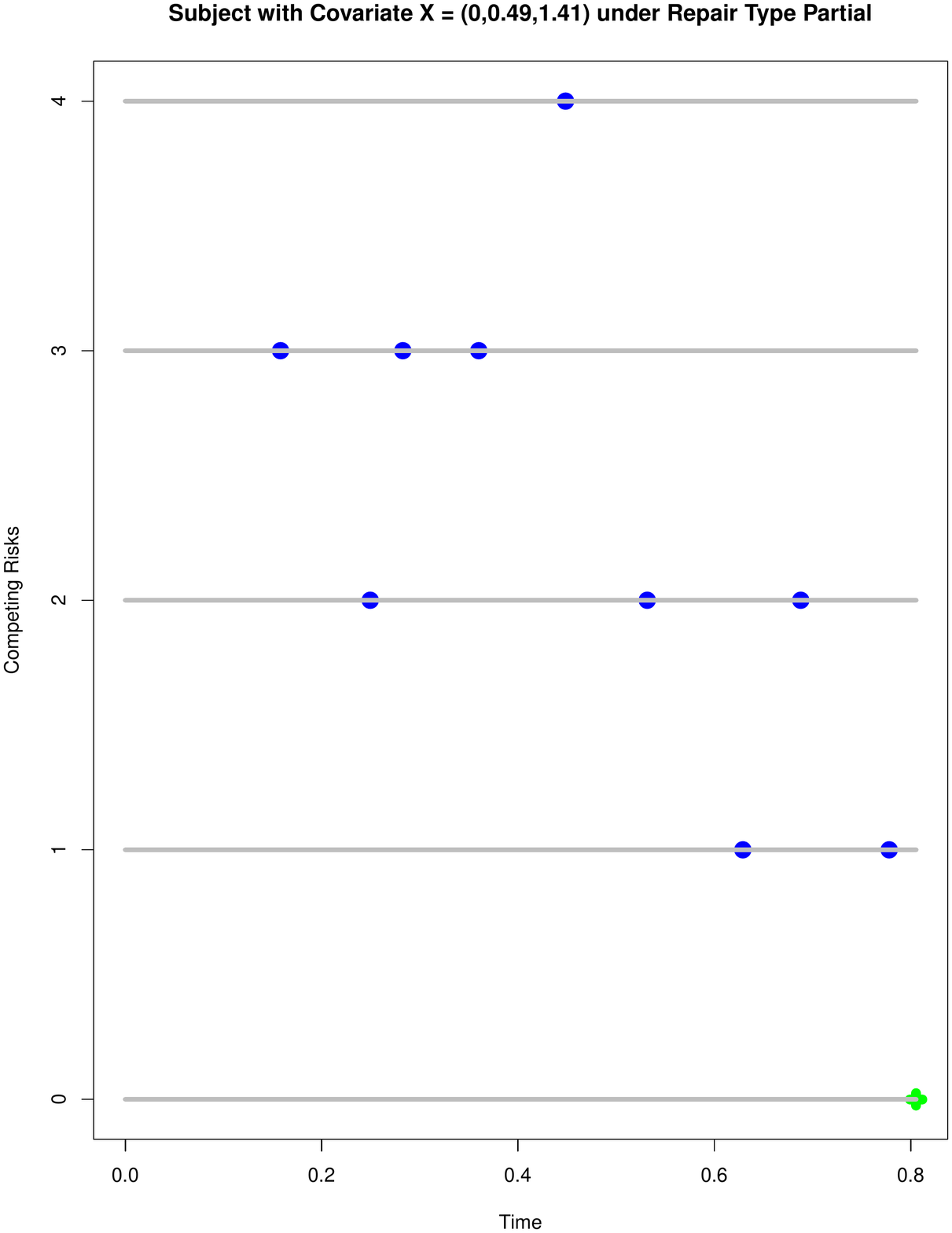}
		\caption{Data History of Unit 0}
		\label{fig:unit0history}
	\end{subfigure}%
	\begin{subfigure}{.5\textwidth}
		\centering
		\includegraphics[width=.8\linewidth, height = 0.25\paperheight]{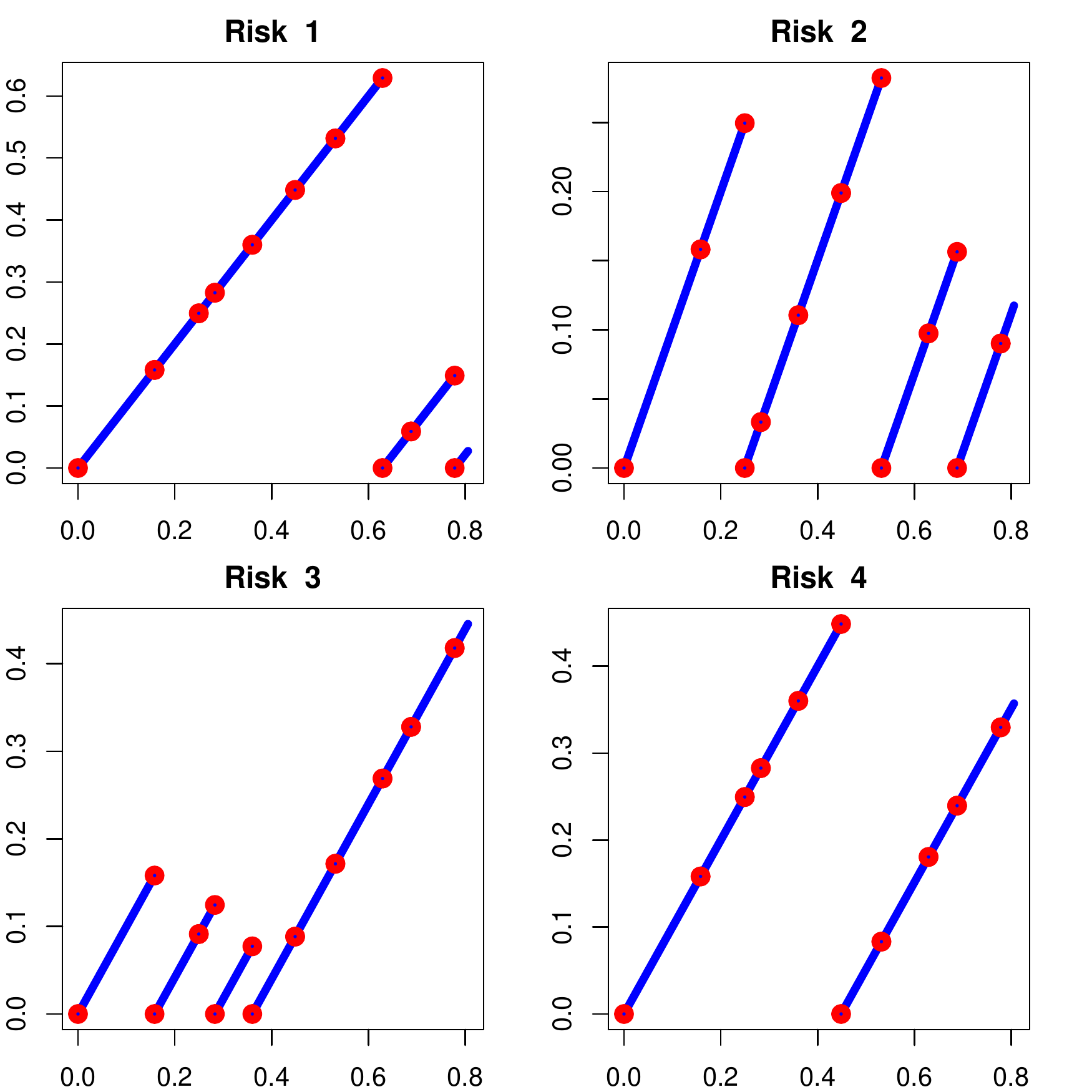}
		\caption{Observed Effective Ages of Unit 0}
		\label{fig: unit0EA}
	\end{subfigure}
	\caption{Prediction for a New Unit: Observed RCRs and Effective Ages of Unit 0}
	\label{fig:unit0data}
\end{figure}

For a new unit 0 that is at-risk by the end of its monitoring time $\tau_0$, we predict its TTTE using the individual history and parameter estimate $\hat{\Theta}$. Although we cannot estimate the value of the frailty variable $Z_0$, we can use its conditional mean as an estimate. Let $\hat{Z}_0$ denote the predicted frailty value of the new unit. The conditional mean of $Z_0$ given history and model parameter is 
\begin{eqnarray}\label{eq: z0}
\Expe[Z_0|\Theta, D_0(\tau_0)] = \frac{\xi + \sum_{q = 1}^{Q}N_{q0}^{\dagger}( \tau_0) + N_{00}^{\dagger}( \tau_0)}{\xi +\sum_{q=1}^{Q}A^{\dagger}_{q0}(\tau_0) + A^{\dagger}_{00}(\tau_0) }
\end{eqnarray}
where  $A^{\dagger}_{q0}(\tau_0)$, and $A^{\dagger}_{00}(\tau_0)$ are the cumulative intensity processes of unit 0. For risk $q$, 
\begin{eqnarray}\label{eq:CondHazardq}
A^{\dagger}_{q0}(s) &=& \int_{0}^{s}Y_0^{\dagger}(v)\rho_{q}(N^{\dagger}_{q0}(v-);\alpha_q)\exp(X_0\beta_{q})
\lambda_{q}(\mathcal{E}_{q0}(v))dv.
\end{eqnarray}
For TE of the new unit, 
\begin{eqnarray}\label{eq:CondHazard0}
A^{\dagger}_{00}(s) &=& \int_{0}^{s}Y_0^{\dagger}(v)\rho_{0}(\textbf{N}^{\dagger}_{0}(v-);\gamma)\exp(X_0\beta_{0})
\lambda_{0}(v)dv.
\end{eqnarray}

In the simulation approach, we estimate $A^{\dagger}_{q0}(\tau_0) $ and $A^{\dagger}_{00}(\tau_0)$. $\hat{Z}^{0}$, or the estimated conditional mean of $Z^{0}$ is then computed following equation (\ref{eq: z0}) by inputting event history of the unit as well as $\hat{A}^{\dagger}_{q0}(\tau_0) $ and $\hat{A}^{\dagger}_{00}(\tau_0)$. 
%
\subsection{Estimating $A_{q0}(s, dw|Z_0,\textbf{z})$ and $A_{00}(dw|Z_0, \textbf{z})$}\label{sec: EstCondHazard}
%
Since effective age processes are incorporated into the RCR sub-models, we utilize the doubly-indexed processes estimating the $\hat{A}^{\dagger}_{q0}(\tau_0) $ and $\hat{A}^{\dagger}_{00}(\tau_0)$. Denote TTTE of unit 0 as $T_0, T_0 > \tau_0$.
We use $D_0(s)$ to denote the data history of unit 0 (see Figure \ref{fig:unit0history}):
\begin{eqnarray}
\textbf{D}_0(s) = \{ \big(Y_0^{\dagger}(s), N_{q0}^{\dagger}(s-), \mathcal{E}_{q0}(s), N_{00}^{\dagger}(s-): s \geq 0, q = 1, 2, \cdots, Q\big), X_{0}, \tau_0\}
\end{eqnarray}

Let $T^{\ast}$ be the maximum observed TTTE of the training observations, and $N_{q.}^{\dagger}(T^{\ast})$ and $N_{0.}^{\dagger}(T^{\ast})$ denote the total number of observed $q_{th}$ RCR occurrences and TE occurrences from the training observations, respectively. $T^{\ast}= \max\{T'_1, T'_2, \cdots, T'_n\}$, where $T'_i = \min(\tau_i, T_i).$ Then $$N_{q.}^{\dagger}(T^{\ast}) = \sum_{i = 1}^{n}N_{qi}^{\dagger}(T'_i); \quad 
N_{0.}^{\dagger}(T^{\ast}) = \sum_{i = 1}^{n}N_{0i}^{\dagger}(T'_i)$$

Values of the Q RCR compensator processes are estimated on the scale of effective ages (see Section \ref{sec:ModelEst}). Conditional on the history of unit 0, we need to estimate its instantaneous probabilities of new event occurrences for each risk. 

To estimate the compensator process values of unit 0 beyond $\tau_0$, we use observed effective ages of risk $q$ and $\hat{\Theta}$ from the training set, where $w_l, l = 1, 2, \cdots, N_{q.}^{\dagger}(T^{\ast}).$
For $q = 1, 2, \cdots, Q$, conditional on $\hat{Z}_0$, we estimate the \textit{generalized at-risk process} of unit 0 as 
\begin{eqnarray}\label{eq: genAtRisk}
\hat{Y}_{q0}(s, w|\hat{Z}_0) = \hat{Z}_0\sum_{j = 1}^{N_0^{\dagger}(\tau_0)} I \big\{w \in \mathcal{E}_{q0} (S_{q0j-1}), \mathcal{E}_{q0}(S_{q0j})\big\} \times 
\frac{\rho_q(\textbf{N}^{\dagger}_0(\mathcal{E}^{-1}_{q0j}(w))-;\hat{\alpha}_q)}{\mathcal{E}^{'}_{q0}(\mathcal{E}^{-1}_{q0j}(w))}
\end{eqnarray}
To simulate a single RCR event, we compute $\hat{A}_{q0}(s, dw_l|\hat{Z}_0,\textbf{z}),$ 
\begin{eqnarray}\label{eq: condProb}
\hat{A}_{q0}(s, dw_l|\hat{Z}_0, \textbf{z}) = \hat{Y}_{q0}(s, w|\hat{Z}_0) \hat{\Lambda}_{q0}(s, dw_l|\textbf{z})
\end{eqnarray}
where $\hat{\Lambda}_{q0}(s, dw_l,|\textbf{z})$ is the Nelson-Aalen type of estimates of baseline hazard we obtain from training data, according to equation (\ref{eq: frailLambdaEst}) and the EM algorithm in Section \ref{sec: EM}. Values of the $\rho_q(.)$ function will update dynamically as each simulated RCR takes place. 

To simulate an event occurrence for the TE, we also estimate the conditional instantaneous probability of TE. Values of $\hat{\Lambda}_{0}(dw|\textbf{z})$ are obtained from training observations as well (see Section \ref{sec:ModelEst}). $\hat{A}_{00}(dw|\hat{Z}_0, \textbf{z})$ is then
\begin{eqnarray}\label{eq: CHazardq}
\hat{A}_{00}(dw|\hat{Z}_0, \textbf{z}) = \hat{Z}_0Y_0^{\dagger}(w)\rho_{0}(\textbf{N}^{\dagger}_{0}(w-);\hat{\gamma})\exp(X_0\hat{\beta}_{0})\hat{\Lambda}_{0}(dw|\textbf{z}) 
\end{eqnarray}

We emphasize that both $\hat{\Lambda}_{0}(dw| \textbf{z}) $ and $\hat{\Lambda}_{q0}(s, dw_l| \textbf{z})$ are functions of estimated frailty and parameter values obtained from training observations. $\hat{A}_{q0}(s, d\textbf{w}| \hat{Z}_0,\textbf{z})$ and $\hat{A}_{00}(dw|\hat{Z}_0,  \textbf{z})$ then combine information from both the training data and individual data history of unit 0.

\subsection{The Simulation Algorithm}\label{sec:algorithm}

  We simulate occurrences of the RCRs and the TE for unit 0 according to the following algorithm:
\begin{enumerate}
	\item Let $\widetilde{T}_0$ be the calendar time. Set $\widetilde{T}_0 = \tau_0$. $ \vec{\mathcal{E}}(\widetilde{T}_0) = \big(\mathcal{E}_{10}(\widetilde{T}_0), \mathcal{E}_{20}(\widetilde{T}_0),\cdots, \mathcal{E}_{Q0}(\widetilde{T}_0) \big)^{t}$ is the vector of effective ages of unit 0 by $\widetilde{T}_0.$ 
	We begin with $k = 0$, the total number of simulated RCRs.
	
	
\item Simulate an occurrence of RCR: 
       \begin{enumerate}
     	\item For $q = 1, 2, \cdots, Q:$
      	      \begin{enumerate}
       	      	\item Create vector 
       	      	$\textbf{w} = \big(w_{k_1}, w_{k_2}, \cdots, w_{k_L}\big)^{t}$ where $w_{k_l}$'s are in ascending order, and  $w_{k_l} > \mathcal{E}_{q0}(\widetilde{T}_0),$ and $\{k_1, k_2, \cdots, k_L\} \subseteq \{1, 2, \cdots, N_{q.}^{\dagger}(T^{\ast})\}$, 
       	      	\item Compute $\hat{A}_{q0}(s, dw_{k_l}|\hat{Z}_0, \textbf{z}), l = 1, 2, \cdots, L$.
       	      	\item For $k_1, k_2,\cdots, k_L, l = 1, 2, \cdots, L$:\\
       	      	\begin{itemize}
       	      		\item 	Generate a Bernoulli random variate $B_{l}$ with success probability $\hat{A}_{q0}(s, dw_{k_l}|\hat{Z}_0,\textbf{z})$:\\
       	      		\newline
       	     	If $B_{l} = 1, t_q = w_{k_l} $, else l = l+1;\\
       	      		\newline
       	      		If $l = L, t_q = w_{k_L}.$
       	      	\end{itemize}
       	      
       	      \end{enumerate}
       	\item $\min t_q = \min (t_1, t_2,\cdots, t_Q).$ Go to step 3.
      \end{enumerate}
\item Simulate an occurence of TE:
        \begin{enumerate}
        	\item Create vector 
        	$\textbf{T} = (T_{k'_1}, T_{k'_2}, \cdots, T_{k'_M})^{t}$, where $T_{k'm}$'s are in ascending order, and $T_{k'm} > \widetilde{T}_0,$ and $\{k'_1, k'_2, \cdots, k'_M\} \subseteq \{1, 2, \cdots, N_{0.}^{\dagger}(T^{\ast})\}.$
        	\item Compute $\hat{A}_0(dT_{k'_m}|\hat{Z}_0,\textbf{z}), m = 1, 2, \cdots M.$
        	\item For $k'_1, k'_2, \cdots, k'_M, m = 1, 2, \cdots, M$:
        	
        	\begin{itemize}
        		\item 	Generate a Bernoulli random variate $B_{k'm}$ with success probability $\hat{A}_{00}(dT_{k'_m}|\hat{Z}_0,\textbf{z})$:\\
        		\newline
        		If $B_{m} = 1, T^{\ast}_0 = T_{k'_m} - \widetilde{T}_0 $, else m = m+1
        		\newline
        		If $m = M, T^{\ast}_0 = T_{k'_M} - \widetilde{T}_0.$
        	\end{itemize}
        \end{enumerate}
        	\item Simulate a single TE path:\\
        \newline
        	 If $T^{\ast}_0 < \min t_q$, 
        		       \begin{enumerate}
        		       	\item Stop.
        		       	\item Update $\widetilde{T}_0 = \widetilde{T}_0 + T^{\ast}_0$.
        		       	\item Update $\vec{\mathcal{E}}(\widetilde{T}_0),$ according to the type of repair (perfect vs. partial).
        		       	\item Set $\hat{T}_0 = \widetilde{T}_0$.
        		       \end{enumerate} 
        		Else:
        		    \begin{enumerate}
        		    	\item Update $\widetilde{T}_0 = \widetilde{T}_0 + T^{\ast}_0,$ . 
        		    	 \item Update $ \vec{\mathcal{E}}(\widetilde{T}_0)$ and $N^{\dagger}_{q0}(\widetilde{T}_0), q = 1, 2, \cdots, Q$, according to the type of repair (perfect vs. partial).
        		    	\item Repeat steps 2, 3 and 4.
        		    \end{enumerate}

        \end{enumerate}
\section{Predictive Accuracy}
To measure predictive performance of the simulation method in Section \ref{sec3: Methods}, we use empirical Brier Score. The Brier Score is a version of expected squared-error loss for evaluating difference between observed and predicted values. Previous works on this topic have detailed description of this measure of predictive accuracy (\cite{graf1999assessment},\cite{gerds2006consistent}, \cite{mauguen2013dynamic}, and \cite{blanche2015quantifying}). The empirical Brier Score is 
\small
\begin{eqnarray}\label{eq:EBS}
\hat{EBS}(v, t) = \frac{1}{\#\{i:Y_i^{\dagger}(v) = 1\}} \sum_{\{i:Y_i^{\dagger}(v) =1\}}  
\hat{w}_i (v+t,\hat{F})\Big[I(T_{i} > v+t) - \hat{P}(T_{i} > v + t|\hat{\Theta}, \mathscr{F}_{v-})\Big]^2
\end{eqnarray}
\normalsize
and the weight $\hat{w}_i(v+t,\hat{F})$ is 
\begin{eqnarray}
\hat{w}_i(v+t, \hat{F}) = \frac{I(v <T_{i} \leq v+t)\{N^{\dagger}_{0i}\big((v,v+t]\big)=1\}}{\hat{F}(T_{i})/\hat{F}(v)}+ \frac{I(T_{i} > v+t)}{\hat{F}(v+t)/\hat{F}(v)} .
\end{eqnarray}
Only observations that are still at-risk at time $v$ will be considered for prediction, and $\hat{F}(.)$ is the Kaplan - Meier estimator of the monitoring time distribution $\tau_i$. When the TE happens in the interval $(v,v+t], N^{\dagger}_{0i}\big((v,v+t]\big)=1.$ 

\section{An Example on Synthetic Data}
In this section, we demonstrate our proposed dynamic prediction method on synthetic datasets. $\hat{\Theta}$ is obtained from a training set of 50 units (see Figure \ref{Manydat}). The true parameter values that we use to simulate the dataset and their estimates are displayed in Table \ref{ParTrue}. We use five different Weibull random variables to create the five baseline hazards of the RCRs and the TE. The baseline hazards are estimated non-parametrically, and are only used for the purpose of generating synthetic datasets. 
\subsection{Parameter Estimates of the Dynamic Joint Model Under Frailty}
The true finite-dimensional parameters and their estimates on the training dataset are displayed in Table \ref{ParTrue}. $\xi$, the frailty variable parameter, is estimated to be 1.506, and the true $\xi$ value is 2.
\begin{table}[H]
	\centering
	\resizebox{\textwidth}{!}{%
	\begin{tabular}{|c|c|c|c|c|}
		\hline
		\textbf{Risk($q$)}  & $\alpha_q$ & $\beta_q$ &$\hat{\alpha}_q$& $\hat{\beta}_q$ \\ \hline
		1   & 0.25& (-0.2, 0.1, 0.30) & -0.106& (-0.34, -0.03, 0.40)\\ \hline   
		2   & 0.2 & (0.3, 0.1, 0.05)  & 0.150 & (-0.07, 0.11, -0.03)\\ \hline
		3   & 0.1 & (0.3, -0.1, 0.40) & 0.103 & (0.05, -0.17, 0.39)  \\ \hline
		4& 0.05 & (0, 1, -0.5) & -0.020 & (-0.18, 0.82, -0.44) \\ \hline	
		\textbf{TE} & $\gamma$ & $\beta_0$ & $\hat{\gamma}$ &$ \hat{\beta_0}$\\ \hline		
		     & (0.1, 0.1, 0.05, 0.5) & (0.3, -0.4, 0.5)& (-0.04, 0.007, 0.038, 0.657) & (0.028, -0.714, 1.076)\\ \hline
	\end{tabular}
	}
	\caption{True Finite-Dimensional Parameters and Estiamtes}
	\label{ParTrue}
\end{table}	

\begin{figure}[H]
	\centering
	\makebox[\linewidth]{%
		\includegraphics[scale = 0.5,width=0.35\paperwidth,height=3in,angle = 0]{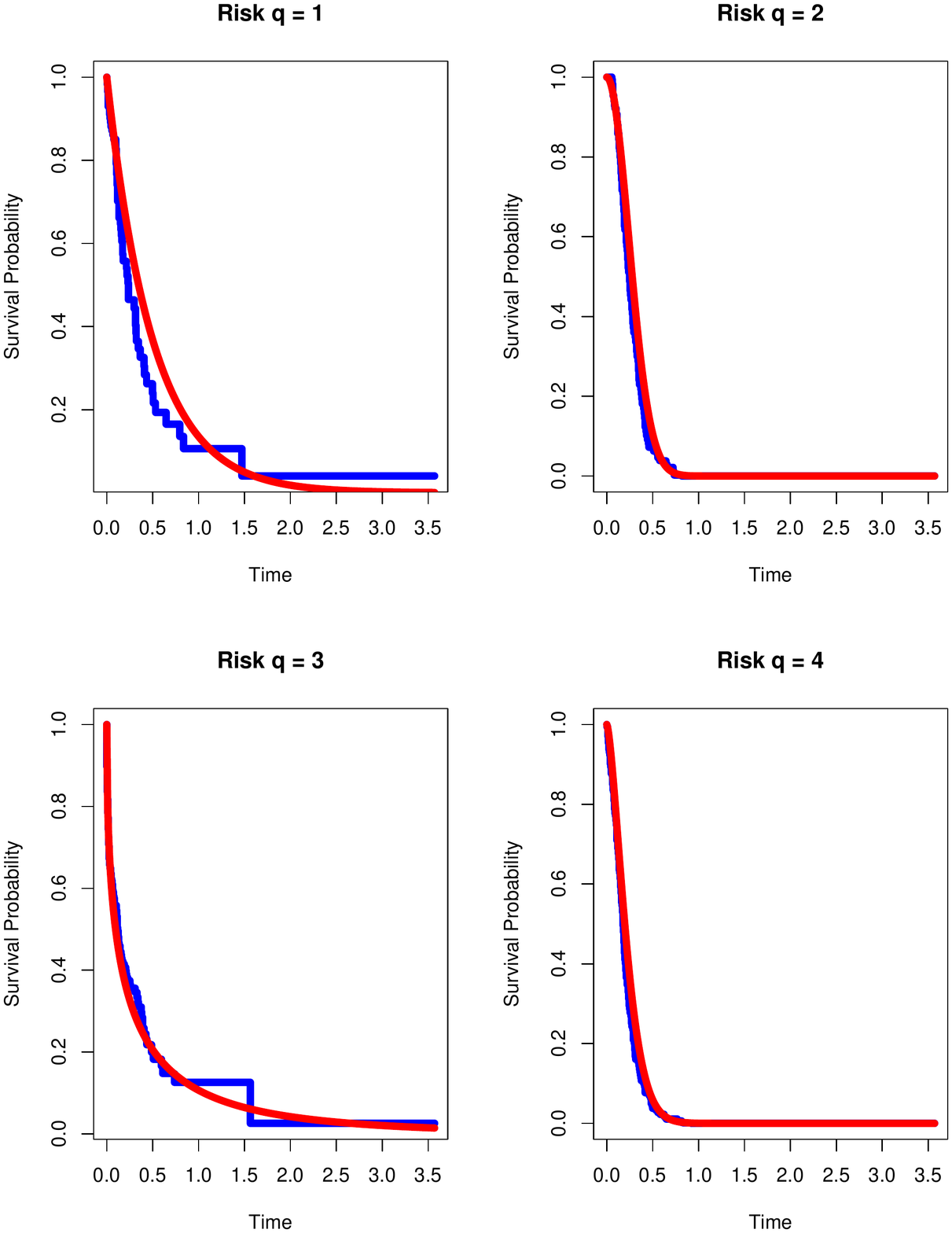}
		\includegraphics[scale = 0.5,width=0.35\paperwidth,height=3in, angle = 0]{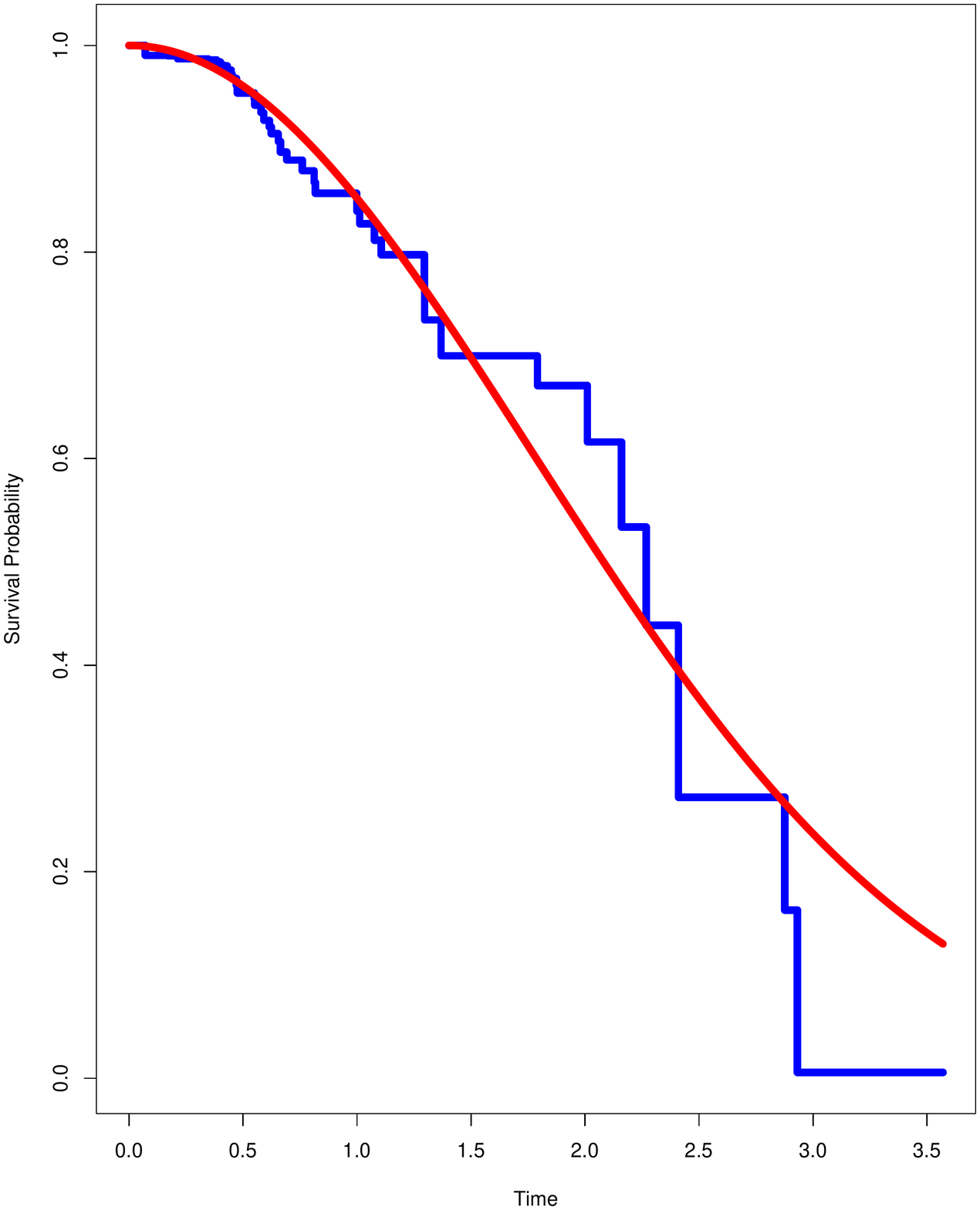}}				
	\caption{\textit{Left}: PLE Survival function of Recurrent Competing Risks; \textit{Right}: PLE Survival function of Terminal Event}
	\label{PLEwithFrail}
\end{figure}

The Product Limit Estimates of the survival functions for the RCRs and TE are displayed in Figure \ref{PLEwithFrail}. The blue wiggly lines are the Kaplan-Meier estimates of the baseline survival functions,, and they are very close to the true survival functions in red. 

\subsection{Predicting Time to TE of a New Unit}
To illustrate our proposed prediction method, we use unit 0 (see Figure (\ref{fig:unit0history})) under partial repair as an example. The algorithm in Section \ref{sec:algorithm} is applicable to the case of perfect repair as well, as one only needs to alter the updating of effective ages in the algorithm. When there are fewer number of RCRs, such as when Q = 2 for monitoring patients who suffer from cardiovascular diseases (see Figure \ref{fig2:dataset2}), the algorithm is easily carried over and simplified. We only demonstrate the case of partial repair in this paper.  

\begin{figure}
	\centering
	\begin{subfigure}{.5\textwidth}
		\centering
		\includegraphics[width=.6\linewidth]{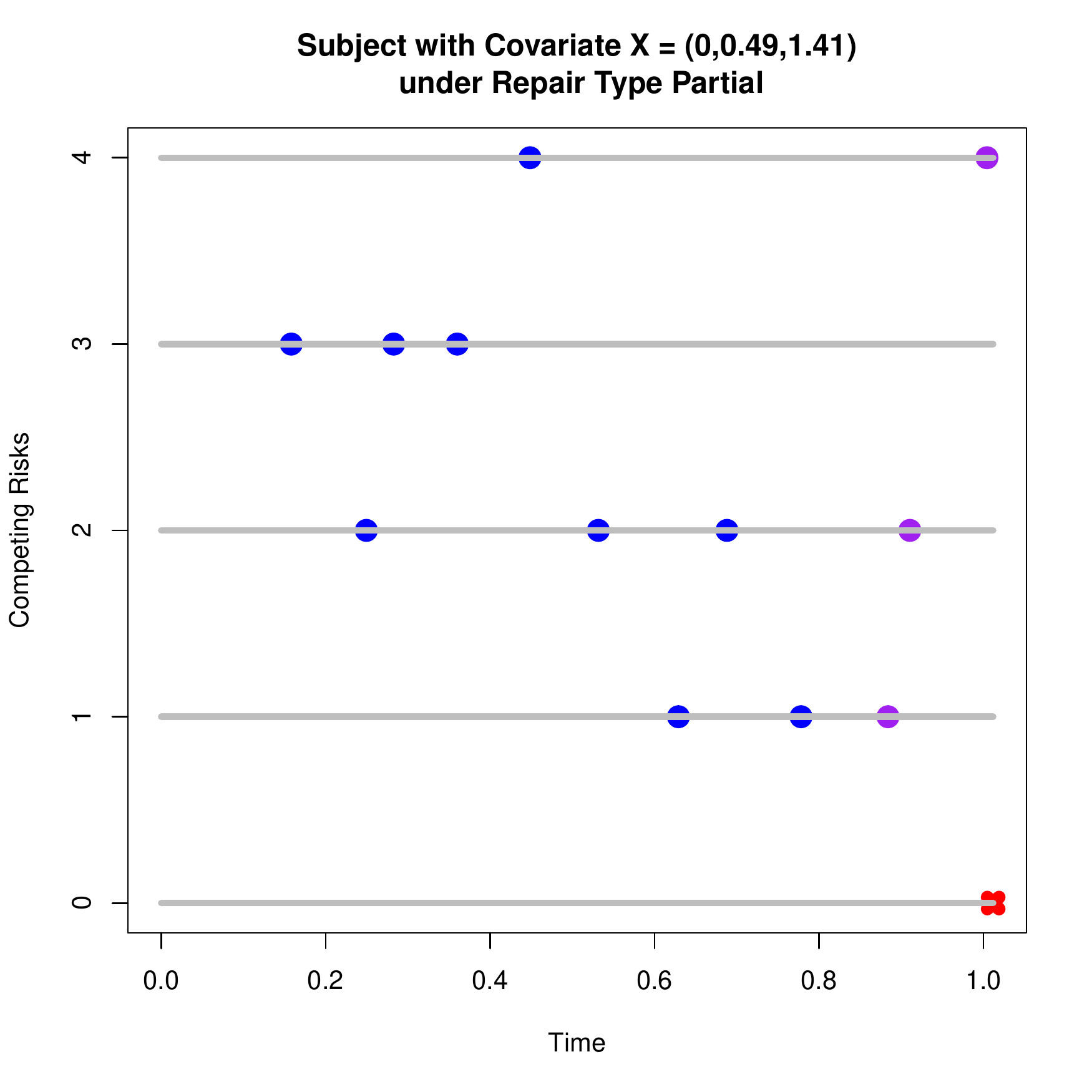}
		\caption{Data History and Simulated RCRs}
		\label{fig:simulevent}
	\end{subfigure}%
	\begin{subfigure}{.6\textwidth}
		\centering
		\includegraphics[width=.5\linewidth]{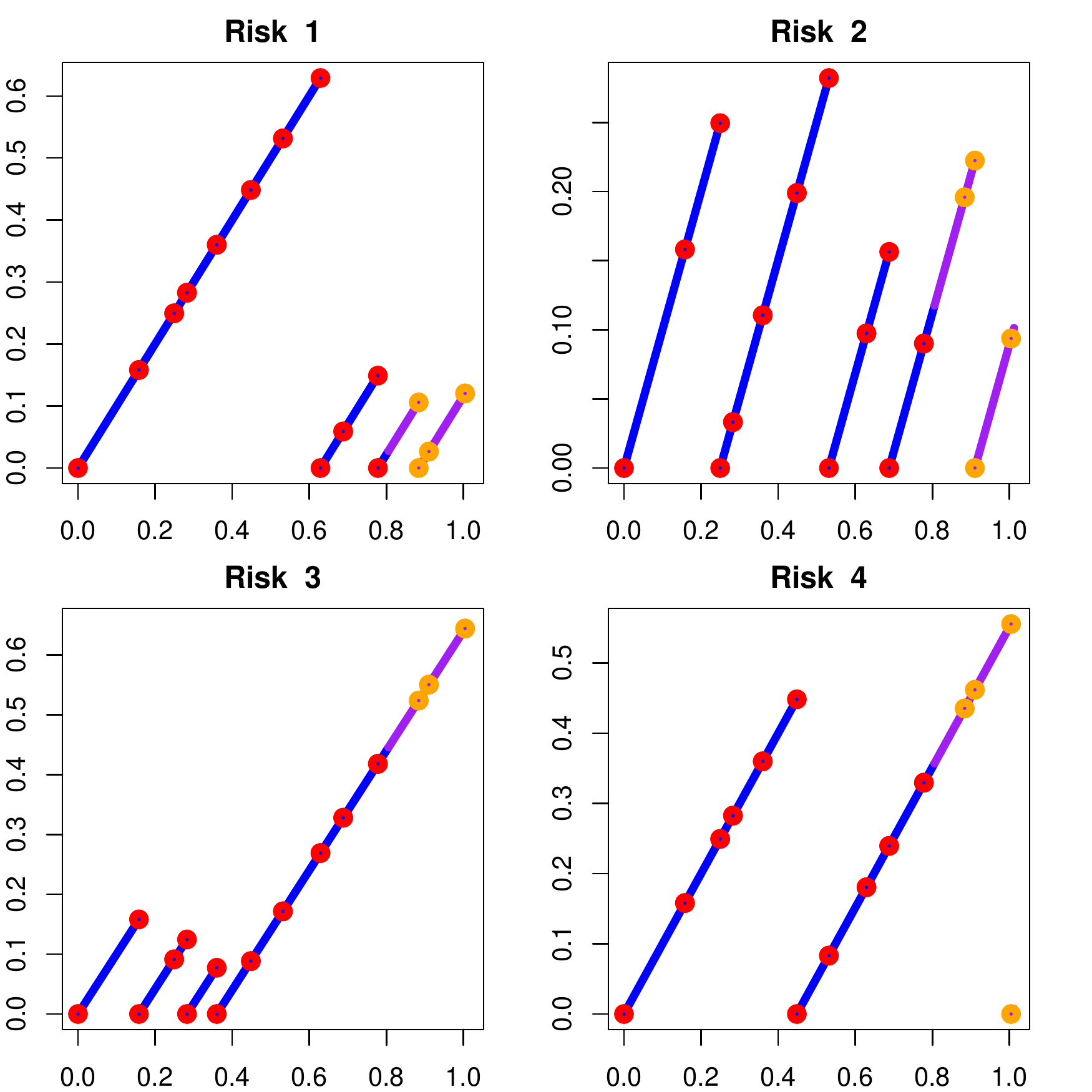}
		\caption{Simulated EAs Under Partial Repair}
		\label{fig: simulEAunit0}
	\end{subfigure}
	\caption{Data History and Predictions of Unit 0 (One Simulated Path)}
	\label{fig:unit0simul}
\end{figure}

For an illustration, predicted RCR occurrences after $\tau_0$ until the occurrence of TE are provided in Figure \ref{fig:unit0simul}. In Figure (\ref{fig: simulEAunit0}), under partial repair, we show simulated effective ages of unit 0 as a result of one single path of simulated TTTE. Solid orange dots represent simulated RCRs, and purple lines represent effective ages. In total, three RCRs are simulated beyond $\tau_0$. In Figure (\ref{fig:simulevent}), observed RCRs of unit 0 and the simulated RCRs are plotted together. Solid purple dots represent simulated RCR event occurrences. No risk 3 event is observed, and each of the other risks is predicted to experience one event until the occurrence of TE. This simulation presents one possible scenario how data accrual of unit 0  happens after $\tau_0$. 

\begin{figure}
	\centering
	\includegraphics[width=0.5\paperwidth,height = 0.25\paperheight]{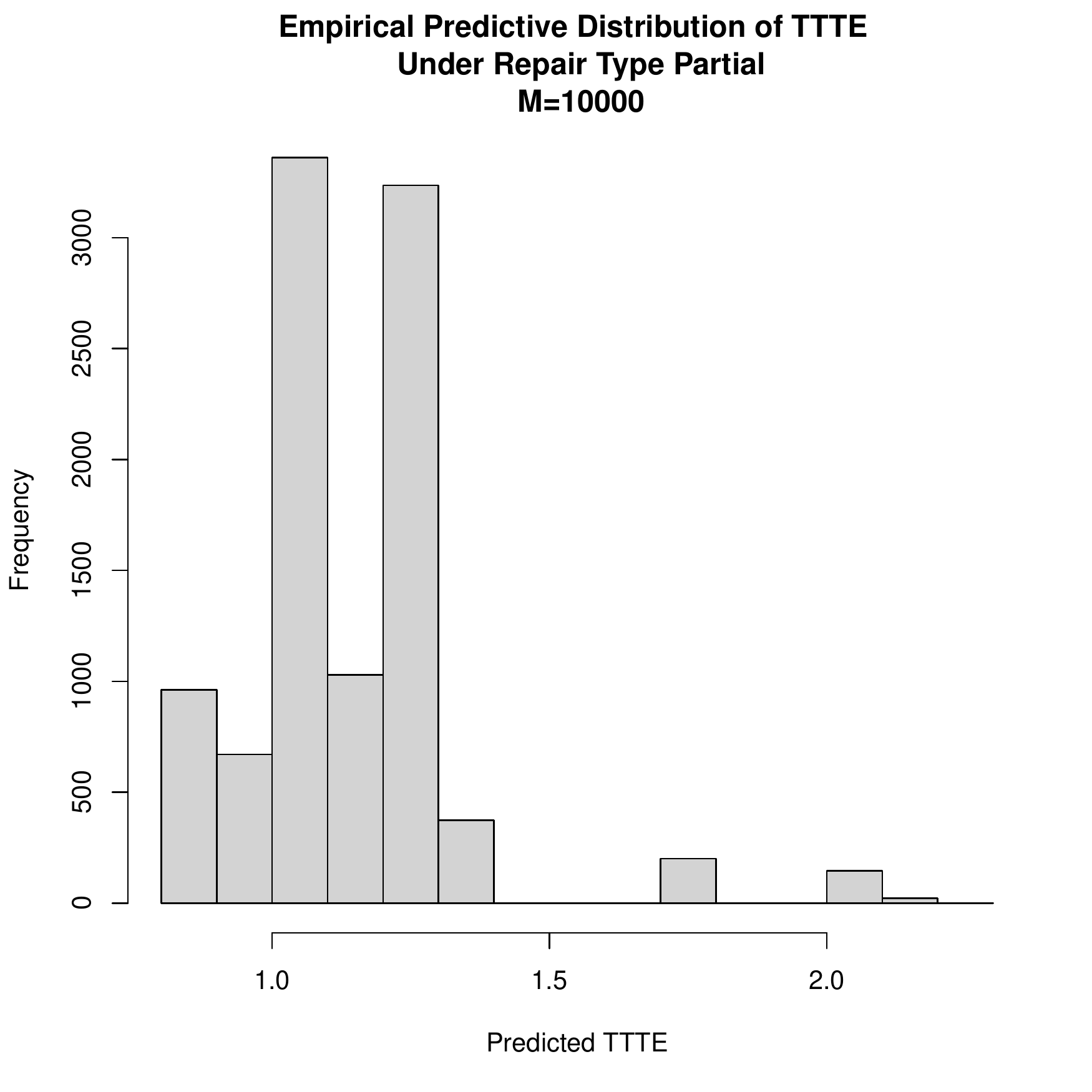}
	\caption{Empirical Predictive Distribution of TTTE (M = 10,000 paths)}
	\label{fig:distT0}
\end{figure}

From a single simulated path of TE, we obtain one realization of predicted TTTE. From a large number of simulated paths, we are able to obtain an empirical predictive distribution of TTTE, denoted by $\hat{T}_0$. In Figure \ref{fig:distT0}, histogram of M = 10,000 paths of the simulated TTTE is shown. 

\subsection{Predicted Survival Probability of the TE at $\tau_0 + s$}
For unit 0, we obtain the predicted survival probability of TE on $(\tau_0, \tau_0+s]$ according to,
\begin{eqnarray}
\hat{P}(T_0 > \tau_0 + s|\hat{\Theta}, \mathscr{F}_{\tau_0-}) = \frac{\sum_{i = 1}^{M}I (\hat{T}_0 > \tau_0 + s)}{M}.
\end{eqnarray}

\begin{figure}
	\centering
	\begin{subfigure}{.5\textwidth}
		\centering
		\includegraphics[width=.8\linewidth, height = 0.25\paperheight]{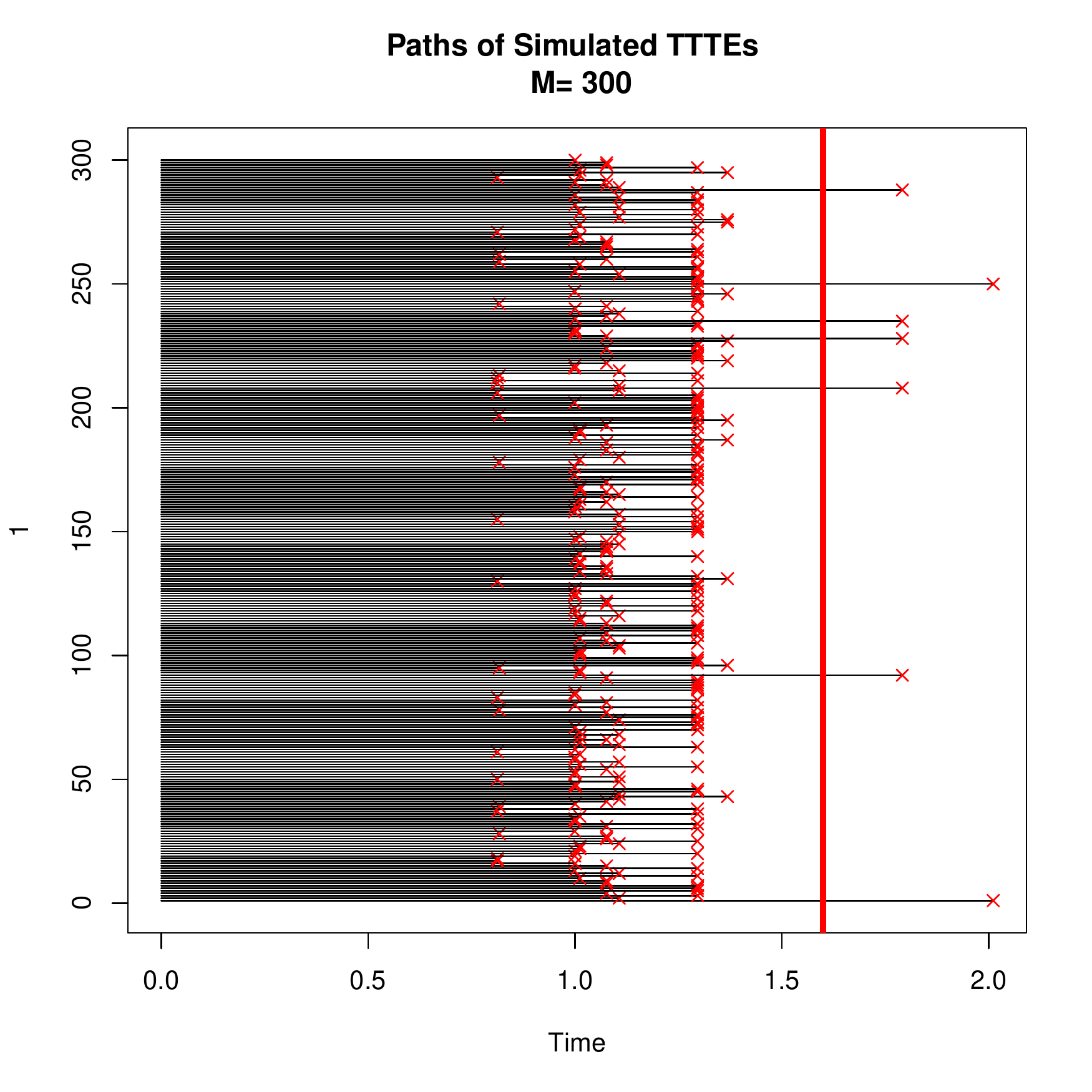}
		\caption{300 Simulated Paths of TTTE}
		\label{fig: manyTEs}
	\end{subfigure}%
	\begin{subfigure}{.5\textwidth}
		\centering
		\includegraphics[width=.8\linewidth, height = 0.25\paperheight]{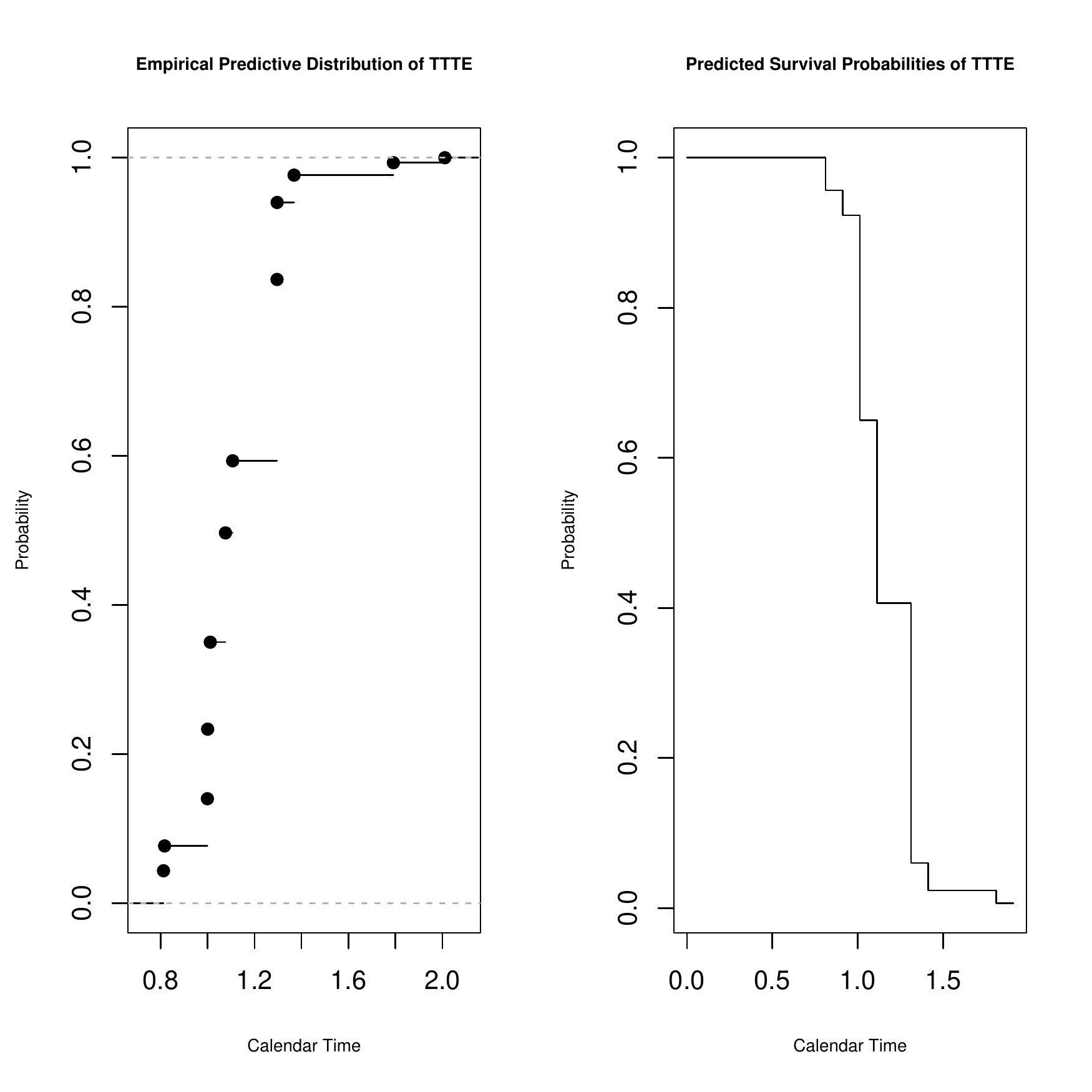}
		\caption{Empirical CDF (\textit{left}) and Predicted Survival Probabilities of TTTE (\textit{right})}
		\label{fig: survprobs}
	\end{subfigure}
	\caption{A Demonstration: Many Simulated Paths of TTTE of Unit 0}
	\label{fig:unit0manypaths}
\end{figure}

If we simulate a large number of TTTE paths, the predicted survival probability of TE at $\tau_0 + s$ is the percentage of them which are greater than $\tau_0 + s$. In Figure (\ref{fig: manyTEs}), we show M = 300 simulated TTTE paths, where the red crosses indicate TE occurrences of unit 0. Lengths of the horizontal lines represent lengths of the simulated TTTE. The predicted survival probability of at time 1.6 is the percentage of red crosses out of M appearing after the vertical redline. Among these M = 300 simulations, there are about 2.3\% of the paths being greater than 1.6. Consequently, the predicted survival probability is about 0.023. In Figure (\ref{fig: survprobs}), predicted survival probabilities can be plotted at different time windows of interest after $\tau_0$ (see the right panel of Figure (\ref{fig: survprobs})). We can also obtain an empirical CDF of TTTE (see left panel of Figure (\ref{fig: survprobs})). 


For M = 10,000 paths, the average and median number of total predicted RCR occurrences per path are 15 and 13, respectively. The average
simulated TTTE is 1.146 (see Table \ref{tab: aveSimulTTTE}). For each risk, we summarize the average and median number of predicted RCR event occurrences per path 
in Table \ref{tab: SimulRCREvent}.
\begin{table}
\centering
\begin{tabular}{ | c | c| } 
\hline
Average Total RCR Occurrences & Average RCR Occurrences After $\tau_0$ \\ 
\hline
24 & 15\\
\hline
\end{tabular}
\caption{Average RCR Event Occurrences of Unit 0 (M = 10,000) Per Path}
\label{tab: SimultotalEvent}
\end{table}
\begin{table}
\centering
\begin{tabular}{|c | c | c| c |c| } 
\hline
 Risks &q = 1 & q = 2 & q = 3 & q =4\\ 
\hline
Average &2.360& 2.220& 9.402& 1.574  \\ 
\hline
Median & 2 & 2 & 7 & 2\\
\hline
\end{tabular}
\caption{Average and Median Simulated RCRs of Unit 0 Per Path (M = 10,000)}
\label{tab: SimulRCREvent}
\end{table}
\begin{table}
\centering
\begin{tabular}{ | c | c |c| } 
\hline
Ave. TTTE & $2.5^{th}$ Percentile  & $97.5^{th}$ Percentile\\ 
\hline
1.146  & 0.812 & 1.791  \\ 
\hline
\end{tabular}
\caption{Summary Statistics of Simulated TTTEs of  Unit 0 (M=10,000)}
\label{tab: aveSimulTTTE}
\end{table}


\subsection{Empirical Brier Score}

\begin{figure}
	\centering
	\includegraphics[width=.8\linewidth, height = 0.25\paperheight]{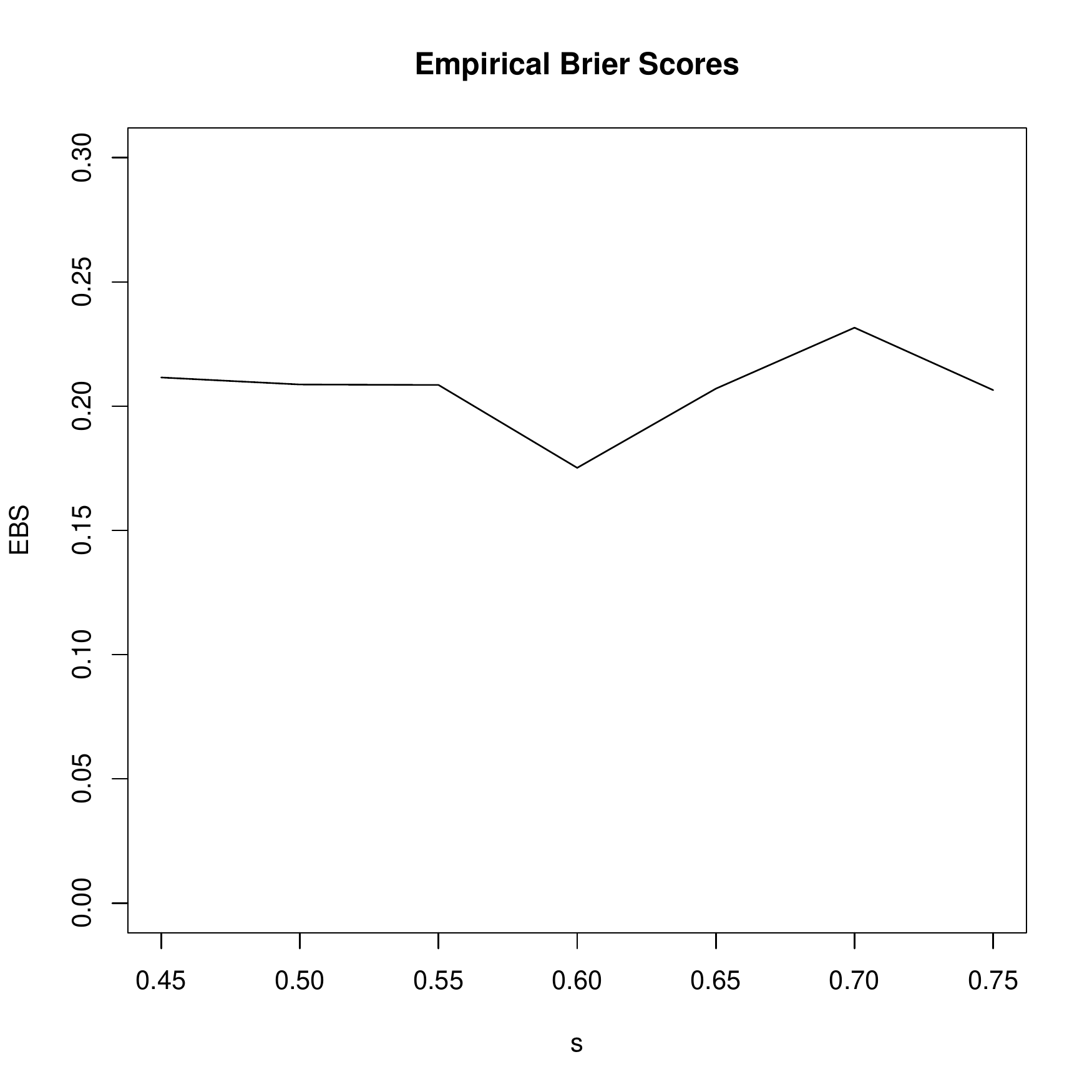}
	\caption{Empirical Brier Scores}
	\label{fig: EEbrier}
\end{figure}

We perform 5-fold cross-validation on a synthetic dataset of 100 observations. The dataset is generated using true model parameter values in Table \ref{ParTrue}. For each of the five iterations, joint dynamic models are trained on $n = 80$ observations, and the prediction method is evaluated on the remaining 20 observations where the observational units are observed up to time $t = 0.8$. We predict survival probabilities $P(T > t + s|\Theta, \mathscr{F}_{t-}) $ for different values of $s$, ranging from 0.45 to 0.75 with 0.05 as the increment. Each predicted probability is computed based on $M = 500$ simulated paths. The empirical Brier scores (see Figure \ref{fig: EEbrier}) at different values of $s$ are observed to be around 0.2. 

\section{Concluding Remarks}

In this paper, we propose a simulation method to predict TTTE dynamically according to a class of joint dynamic models of RCRs and TE (\cite{liu2015dynamic}). By estimating the conditional instantaneous probabilities of Q RCRs and TE, the simulation method combines individual data history of a new unit and parameter estimates obtained on a training set to make \textit{personalized} prediction of TTTE. The method provides an empirical predictive distribution of TTTE by simulating a large number of paths of TTTE. For each simulated path, RCRs are simulated leading up to the TE. The RCRs and TE are generated dynamically as each simulated RCR occurrence increases knowledge we obtain about a unit and hence updates the risks of new RCRs and the TE. For a new unit that is observed by some time of interest $t$, the predicted survival probability of TE at $t+s$ is also obtained, where $s$ is some time window of interest. Important quantities regarding the Q RCRs can be summarized from a large number of simulated TTTE paths. Predictive accuracy of the proposed method is evaluated using empirical Brier score. 
Although the proposed prediction method uses parameter estimates from a particular class of joint dynamic models, the simulation approach to dynamically predicting TTTE can be adopted by other joint dynamic models. As data structures calling for joint dynamic modeling can potentially become much more complicated than what has been considered in this project, our ongoing work is to apply and extend the simulation method in predicting TTTE to adapt to more challenging data situations.
\bibliographystyle{plain}
\bibliography{RCRTEPred}


\end{document}